\documentclass[pdflatex, iicol, sn-nature]{sn-jnl}

\usepackage{graphicx}%
\usepackage{multirow}%
\usepackage{amsmath,amssymb,amsfonts}%
\usepackage{amsthm}%
\usepackage{mathrsfs}%
\usepackage[title]{appendix}%
\usepackage{xcolor}%
\usepackage{textcomp}%
\usepackage{manyfoot}%
\usepackage{booktabs}%
\usepackage{algorithm}%
\usepackage{algorithmicx}%
\usepackage{algpseudocode}%
\usepackage{listings}%
\usepackage{siunitx}%
\usepackage{subcaption} 
\usepackage{csquotes}
\usepackage{braket}
\usepackage{bbold}
\usepackage{physics}
\usepackage{booktabs}
\usepackage{tikz}  
\usetikzlibrary{tikzmark}
\usepackage{textcomp}
\usepackage{enumitem}

\theoremstyle{thmstyleone}%
\theoremstyle{thmstyletwo}%

\theoremstyle{thmstylethree}%

\begin{document}

\title[]{Highly indistinguishable photons from a tin-vacancy spin qubit in diamond}

\author[1,2]{\fnm{Dennis} \sur{Herrmann}}
\equalcont{These authors contributed equally to this work.}

\author[1,2]{\fnm{Robert} \sur{Morsch-Golsong}}
\equalcont{These authors contributed equally to this work.}

\author[1,2]{\fnm{Tobias} \sur{Bauer}}
\equalcont{These authors contributed equally to this work.}

\author[1,2]{\fnm{Marlon} \sur{Schäfer}}
\equalcont{These authors contributed equally to this work.}

\author[1,2]{\fnm{David} \sur{Lindler}}
\equalcont{These authors contributed equally to this work.}

\author[1,2]{\fnm{Linus} \sur{Ehre}}

\author[3]{\fnm{Peter} \sur{van Loock}}

\author[4]{\fnm{Matthew} \sur{Markham}}

\author[4]{\fnm{Nicola} \sur{Palmer}}\

\author[5]{\fnm{Soumen} \sur{Mandal}}

\author[5]{\fnm{Oliver} \sur{Williams}}

\author*[1,2]{\fnm{Christoph} \sur{Becher}}\email{christoph.becher@physik.uni-saarland.de}

\affil[1]{\orgdiv{Fachrichtung Physik}, \orgname{Universität des Saarlandes}, \orgaddress{\street{Campus E2.6}, \city{Saarbrücken}, \postcode{66123}, \country{Germany}}}

\affil[2]{\orgdiv{Zentrum für Quantentechnologien (QuTe)}, \orgname{Universität des Saarlandes}, \orgaddress{\street{Campus}, \city{Saarbrücken}, \postcode{66123}, \country{Germany}}}

\affil[3]{\orgdiv{Institut für Physik}, \orgname{Johannes Gutenberg-Universität Mainz}, \orgaddress{\street{Staudingerweg 7}, \city{Mainz}, \postcode{55128}, \country{Germany}}}

\affil[4]{\orgdiv{Global Innovation Centre}, \orgname{Element Six (UK) Ltd.}, \orgaddress{\street{Fermi Avenue, Harwell Campus}, \city{Didcot}, \postcode{OX11 0QR}, \country{United Kingdom}}}

\affil[5]{\orgdiv{School of Physics and Astronomy}, \orgname{Cardiff University}, \orgaddress{\street{The Parade}, \city{Cardiff}, \postcode{CF24 3AA}, \country{United Kingdom}}}

\abstract{
Quantum networks \cite{Wehner2018,Wei2022} promise secure communication, distributed sensing and modular quantum computing by interconnecting distant quantum nodes through photonic links. Extending such networks beyond metropolitan distances requires quantum repeaters \cite{Briegel1998, Azuma2015} to overcome the exponential attenuation of photons in optical fiber. Across all architectures, a key requirement is the indistinguishability of single photons, which directly impacts the fidelity of photonic operations based on two-photon interference, such as Bell-state measurements \cite{Simon2003, Rohde2006, Kambs2018} and fusion gates \cite{Chan2025,Thomas2024}. Here, we demonstrate generation of highly indistinguishable single photons from a coherently excited tin-vacancy center in diamond \cite{Debroux2021,Trusheim2020,Arjona2022,Gorlitz2022,Kuruma2021}, achieving raw Hong–Ou–Mandel \cite{Hong1987} interference visibilities exceeding 0.95. By separating intrinsic emitter properties from technical imperfections, we show that decoherence plays a negligible role and that the remaining limitations are predominantly technical in nature, arriving at an intrinsic indistinguishability of up to 0.999. We further show that quantum frequency conversion \cite{kumar1990,Brevoord2025,Schaefer25} to the telecom C-band preserves the photon indistinguishability. In combination with the long-lived electron \cite{Karapatzakis2024} and nuclear \cite{Resch2026} spin coherence times, these results establish tin-vacancy centers in diamond as a competitive platform for long-distance quantum networks and photonic quantum information processing \cite{Azuma2015,Ruf2021,Couteau2023}. We further substantiate this potential through Monte Carlo simulations of a quantum-repeater link, demonstrating that the SnV-center platform surpasses the bound set by direct transmission.
}

\maketitle

\section*{Introduction}\label{sec1}
The typical operating principle of a quantum repeater \cite{Briegel1998} is to divide a communication link into short segments, and to generate and distribute entanglement among the nodes. A defining technical bottleneck across repeater architectures is the fidelity of the operations that connect elementary hardware \cite{Kamin2023}. In memory-based quantum repeaters, entanglement swapping \cite{Pan1998} is implemented by a Bell-state measurement \cite{Cabrillo1999, Simon2003, Kiraz2004} that relies on two-photon interference and was shown to create entanglement of remote quantum memories \cite{Bernien2013, Moehring2007, Hofmann2012, Delteil2016} In measurement-based, all-photonic architectures \cite{Munro2012, Azuma2015}, large-scale, higher-dimensional photonic cluster states \cite{Lindner2009, Thomas2022, Cogan2023} are required, which can be achieved by entangling gates \cite{Economou2010} or fusion \cite{Thomas2024, Istrati2020} of smaller, elementary cluster states. While the memory-based setting requires long memory coherence times and both approaches benefit from high photon creation and detection efficiencies, a general relevant figure of merit is the indistinguishability of single photons. In particular, the visibility of Hong-Ou-Mandel (HOM) interference \cite{Hong1987} directly constrains the fidelity of photonic Bell-state measurements \cite{Simon2003, Rohde2006, Kambs2018} and fusion gates \cite{Chan2025,Thomas2024}, and thus the entanglement fidelity and attainable secret-key fraction in a network \cite{Kamin2023}. In another context, single photon sources with near-unity indistinguishability are a cornerstone for fusion-based photonic quantum computing \cite{Bartolucci2023, Chan2025}.\\
To meet the stringent hardware requirements, diamond-based emitters with embedded quantum memories \cite{Wei2022} have emerged as a leading platform. Nitrogen-vacancy centers have played a pioneering role, enabling the first demonstrations of spin–photon entanglement and heralded entanglement between remote solid-state spin qubits \cite{Bernien2013,Stolk2024}. Building on this progress, group-IV color centers have enabled key experiments such as memory-enhanced quantum communication \cite{Bhaskar2020}, entanglement between remote quantum memory nodes \cite{Knaut2024,Waas2026} as well as first demonstrations of distributed quantum sensing \cite{Stas2026} and blind quantum computing \cite{Wei2025}.
More recently, tin-vacancy (SnV) centers in diamond are attracting significant interest due to their outstanding optical properties, including excellent single-photon purity, high spectral stability, and Fourier-limited linewidth \cite{Gorlitz2020,Trusheim2020,Gorlitz2022}, as well as their coherent spin-photon interface \cite{Rugar2021} with very long electronic spin (up to 10 ms \cite{Karapatzakis2024}) and nuclear spin (more than a second \cite{Resch2026}) lifetimes. However, the two-photon interference visibilities reported so far \cite{Arjona2022, Bushmakin2025} have lagged behind those achieved in competing platforms such as semiconductor quantum dots \cite{Tomm2021, Somaschi2016}, neutral atoms \cite{van_leent_entangling_2022}, and trapped ions \cite{Meraner2020}.\\
Here, we demonstrate highly indistinguishable, single photons from SnV centers in diamond, achieving raw two-photon interference visibilities which place them among these leading platforms. By comprehensively modeling the resonant excitation and emission process we are able to factor out technical limitations and arrive at very high intrinsic photon indistinguishability. Eventually we simulate how the performance of our optical interface impacts the fidelity and rate of entanglement generation across a quantum repeater link, defining the requirements for quantum network operation. Importantly, the results demonstrated here, in combination with earlier achievements on spin memory and interfaces, show that SnV centers already fulfill these requirements.

\section{Resonant Excitation of Single Tin-Vacancy Centers}\label{sec2}
\label{par_resex}
\begin{figure*}[htbp]
  \centering

  \begin{subfigure}[t]{\textwidth}
    \centering
    \includegraphics[width=\linewidth]{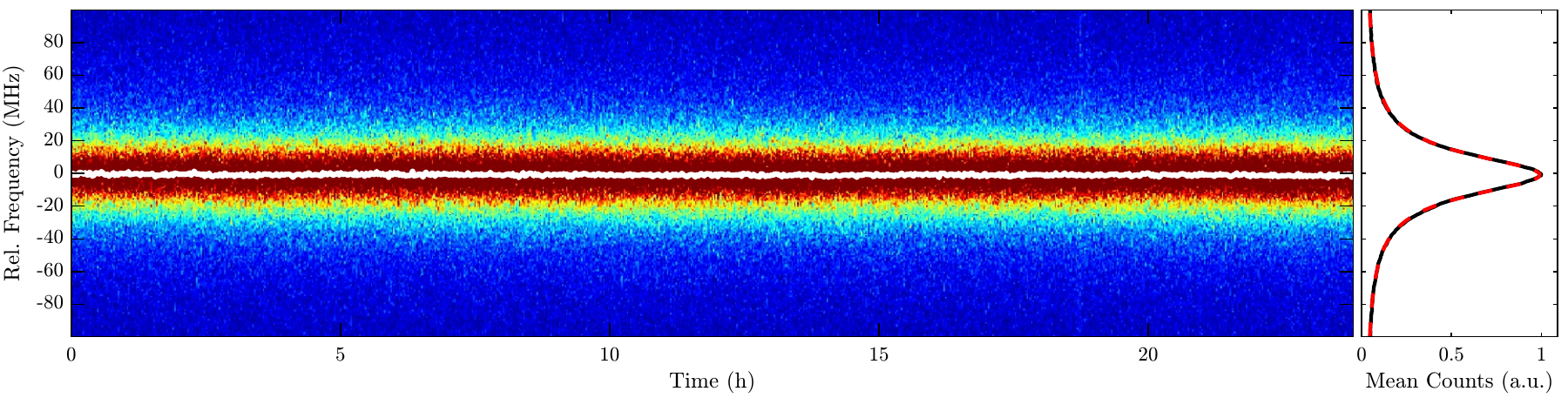} 
    \caption{}
    \label{fig1:PLE}
  \end{subfigure}

\begin{subfigure}[t]{0.32\textwidth}
  \centering  \includegraphics[width=\linewidth]{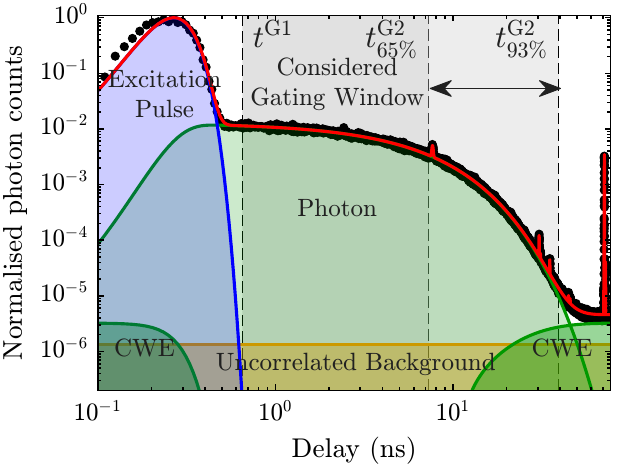}
  \caption{}
  \label{fig1:TCSPC}
\end{subfigure}\hfill
\begin{subfigure}[t]{0.32\textwidth}
  \centering
  \includegraphics[width=\linewidth]{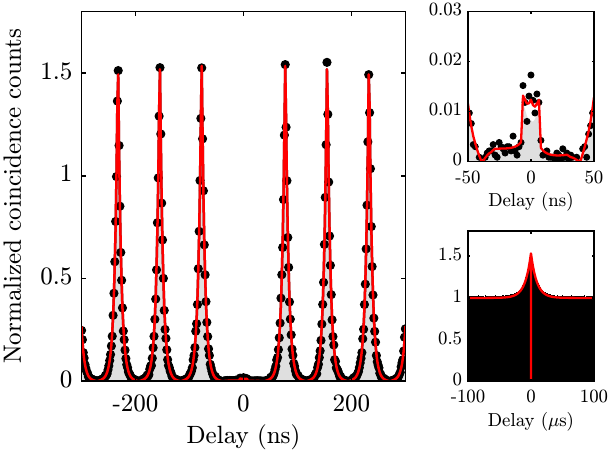}
  \caption{}
  \label{fig1:g2}
\end{subfigure}\hfill
\begin{subfigure}[t]{0.32\textwidth}
  \centering
  \includegraphics[width=\linewidth]{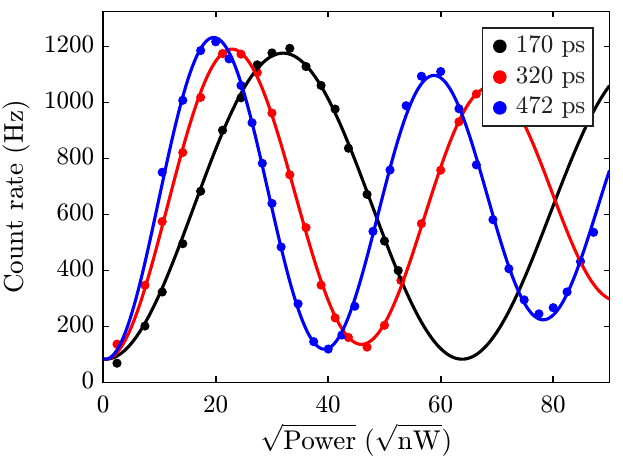}
  \caption{}
  \label{fig1:Rabi}
\end{subfigure}
\caption{\textbf{a}, Photoluminescence excitation spectrum of a single SnV center recorded over 24\,h. The average of all individual scans (solid black line) is fitted with a Lorentzian function (red dashed line), yielding a slightly power-broadened linewidth of $30.8(2)$\,MHz. This value agrees with the linewidth obtained from a single scan, $30.7(4)$\,MHz, indicating negligible spectral drift over the entire measurement duration. Furthermore the linewidth remains close to the Fourier-transform-limited value of $28.9(2)$\,MHz expected from the measured excited-state lifetime of 5.68\,ns. The white trace indicates the central peak position for each individual scan; the observed drift remains below the 2\,MHz uncertainty of the wavemeter. \textbf{b}, Time-correlated single-photon counting histogram for one detection channel during the $g^{(2)}$ measurement. The measured data is fitted with a sum of four relevant contributions (red line), stemming from residual excitation light (blue), emitted photons (light green), and an uncorrelated background (orange), as well as leakage-induced excitation and subsequent single photon emission (CWE, dark green) which appears only after relaxation to the ground state. The applied gating window (gray) excludes the excitation pulse ($t^{\mathrm{G1}} = 0.66\,\mathrm{ns}$) and extends to $t^{\mathrm{G2}}_{93\%} = 38.75\,\mathrm{ns}$, capturing 93\,\% of the photon wave packet. A shorter window ($t^{\mathrm{G2}}_{65\%} = 7.3\,\mathrm{ns}$) is used to suppress multi-path reflections. \textbf{c}, Single photon purity measurement using the 93\% time-gating window, yielding $g^{(2)}_{\mathrm{raw},93\%}(0) = 0.024\pm0.002$. The grey shaded area represents the expected coincidence distribution for perfectly indistinguishable emitter photons, taking into account the measured background noise. \textbf{d}, Power-dependent Rabi-oscillations for different excitation pulse lengths.}
\label{fig1}
\end{figure*}

In a confocal microscope setup, kept at 1.75\,K, SnV centers in an ion-implanted, HPHT-annealed bulk diamond sample (see Methods) are coherently excited via resonant driving of the C transition at 619.248\,nm. Single photons are generated by pulsed coherent excitation. In order to achieve a high coherent population inversion and to minimize multi-photon emission, short $\pi$-pulses with pulse lengths of approximately  170\,ps and pulse extinction ratios of more than 6 orders of magnitude (see Extended Data Figure \ref{EDF:Pulses}) are generated from a cw laser source using two cascaded electro-optic modulators (EOM). Figure \ref{fig1:Rabi} shows the corresponding Rabi oscillations, demonstrating coherent optical control of the emitter. In order to verify the fidelity of the coherent population inversion, we performed photon number state coherence measurements \cite{Loredo2019}, from which we extract an excited-state preparation fidelity of $p_1 = 0.983 _{-0.012}^{+0.013}$ (see Supplementary Information). The photon fluorescence is also detected on the C-transition, necessitating a high extinction of the excitation laser in the detection path. This is realized by a cross-polarization setup \cite{Kuhlmann2013} that allows for polarization-selective filtering exceeding 6 orders of magnitude (see Extended Data Figure \ref{EDF:ER_cross}). 
\\
Solid-state emitters are generally susceptible to fluctuations in the local charge environment, which induce time-dependent Stark shifts of the optical transition. Inversion-symmetric group-IV color centers, such as SnV centers, intrinsically suppress first-order sensitivity to electric-field noise, providing enhanced spectral stability. We complement this intrinsic robustness by actively stabilizing the charge state using an additional laser at 445\,nm \cite{Gorlitz2022}, resulting in excellent long-term stability of the emission line close to the Fourier-limit, as evidenced by the photoluminescence excitation spectrum in Fig. \ref{fig1:PLE}. The integrated linewidth of $30.4_{-3.1}^{+3.6}$\,MHz measured over 24\,h is close to the lifetime-limited value of 28.9\,MHz set by the emitter lifetime of 5.68\,ns.
\\
All of the following experiments were performed with the same single SnV color center. This particular emitter has a lifetime of 5.46\,ns and exhibits spectral jumps on the order of one linewidth (see Extended Data Figure \ref{EDF:PLE_RE1Longterm}). This occurs on a timescale of a few seconds, far slower than the experimental repetition rate, and therefore has negligible impact on the experimental results.
\\
To assess the single photon purity we measure the second-order correlation function  $g^{(2)}(\tau)$ in a Hanbury-Brown-Twiss (HBT) setup, using superconducting-detectors (SNSPD). In a first step, we identify the weights of the relevant noise contributions by fitting the time-correlated single photon counting (TCSPC) data with a four-component model (red line), as shown in Fig. \ref{fig1:TCSPC}. Here, residuals of the reflected excitation pulse (not fully suppressed by the cross-polarization filtering) dominate the exponentially decaying single-photon signal within the first 0.66\,ns, while residual reflections in the setup and at the diamond surface introduce additional sharp features. The finite pulse extinction ratio of the EOMs leads to a weak cw component of the excitation laser inducing two noise contributions. First, some of the excitation laser light is reflected by the sample, which, together with the detector’s dark count noise, results in an uncorrelated background. Second, this light results in a continuous excitation of the emitter, which competes with the pulsed excitation and causes an additional single-photon background.
\\
In order to improve the signal-to-background ratio (SBR), we time-gate the detected photons. The start of the detection window is set to $t^\mathrm{G1}=0.66$\,ns where the excitation pulse contribution reaches the level of the uncorrelated background.
\\
For the end of the detection window, we focus on two cases. In the first, we set the cutoff to $t^\mathrm{G2}_{93\%} = \mathrm{38.75\,ns}$, resulting in a SBR of 53. This corresponds to a gating window equal to half the repetition period, preventing overlap between neighboring peaks in the correlation measurements while retaining 93\,\% of the single-photon wave packet. In the second case, we consider $t^\mathrm{G2}_{65\%}=\mathrm{7.3\,ns}$ to discard all background features caused by reflection, and thus improve the SBR to 637 while still preserving 65\,\% of the photon wave packet.
\\
Fig. \ref{fig1:g2} shows the $g^{(2)}_\mathrm{raw}(\tau)$-function for the first of the aforementioned time-gating windows. Calculating the raw single photon purity as the ratio of the central peak area to the area of the outer peaks far beyond the photon bunching timescale, we obtain $g^{(2)}_\mathrm{raw,93\%}(0)=0.024\pm0.002$ enhanced to $g^{(2)}_\mathrm{raw,65\%}(0)=0.007\pm0.002$ when applying the second gatting window. 
\\
To identify the factors that limit the measured single-photon purity, we model the correlation data by incorporating the individual noise contributions to the auto-correlation function. These contributions are obtained directly from the analytic terms fitted to the TCSPC data and added to the signal-correlation with weights strictly fixed by the corresponding SBR determined in the TCSPC fit under the applied gating conditions. This leaves the noise-free value $g^{(2)}_{\mathrm{bc}}(0)$ as the only free fit parameter (see Methods).
The fit (Fig. \ref{fig1:g2}) shows excellent agreement with the data, confirming that residual reflections of the excitation pulse predominantly limit the raw single photon purity. Correcting for these background contributions yields a background-corrected value of $g^{(2)}_{\mathrm{bc},93\%}(0)=\mathrm{0.004\pm0.002}$ ($g^{(2)}_{\mathrm{bc},65\%}(0)=\mathrm{0.005\pm0.002}$). We note that this high single-photon purity does not contradict the limits derived in Ref. \cite{Fischer2016}, since multi-photon emission occurs predominately within the finite duration of the excitation pulse \cite{Giorgino2025} and is therefore excluded by the temporal gating.

\vspace{2em}

\section{Indistinguishable Single Photons from a Resonantly Excited Tin-Vacancy Center}
In order to investigate the photon indistinguishability of SnV centers, we perform Hong-Ou-Mandel (HOM) measurements on consecutively emitted photons which are overlapped in an unbalanced fiber-based Mach-Zehnder-Interferometer (MZI) with the path-difference matched to the repetition rate $T_{\mathrm{rep}}$. Two-photon interference occurs whenever two photons enter the second non-polarizing beamsplitter of the MZI through different ports, leading to photon coalescence described by the HOM effect \cite{Hong1987} (see Methods).\\ 
Figure \ref{fig2} shows the measured correlations for parallel and orthogonal polarization settings. The red line is obtained by fitting a model to the data that takes into account how the coincidence statistics is affected by intrinsic photonic (pure dephasing, spectral diffusion) and technical (noise, polarization mismatch) sources of imperfection. The model is derived by propagating the temporal gating throughout the formalism proposed in Ref. \cite{Kambs2018} as well as incorporating the independently determined background contributions from the TCSPC measurements (see Methods). This results in a fit function with pure dephasing, spectral diffusion and the polarization mismatch serving as free parameters.  \\
In a first step, the raw HOM visibility is determined from this function fitted to the data by comparing the area of the central peak for both polarization settings, which yields
\(V_\mathrm{HOM}^\mathrm{raw,93\%} = 0.949_{-0.008}^{+0.007}\) 
(\(V_\mathrm{HOM}^\mathrm{raw,65\%} = 0.973_{-0.008}^{+0.008}\)).\\
Additionally, our detailed model accurately reproduces the relative coincidence peak heights (see Supplementary Information), eliminating the need for a reference measurement with orthogonal input polarization setting. Comparing the central peak with the far-delayed peaks, which represent temporally orthogonal photon states, we extract $V_\mathrm{HOM}^\mathrm{raw,93\%} = 0.950_{-0.008}^{+0.006}$
(\(V_\mathrm{HOM}^\mathrm{raw,65\%} = 0.974_{-0.009}^{+0.007}\)), in excellent agreement with the results obtained using the first method. Statistical uncertainties are given as 95\% highest-posterior-density credible intervals derived from the Bayesian posterior distribution (see Supplementary Information). As a third and independent analysis, Fig. \ref{fig3} shows the raw HOM visibility extracted directly from the coincidence histograms after applying a temporal gating of varying width, again yielding values in agreement with the previous evaluations. Longer delays in the MZI up to 155\,ns, yield equally high visibilities.\\
The excellent agreement of the TCSPC fit in Figure \ref{fig1:TCSPC} and the HOM fits in Figure \ref{fig2:hom_vis77} prove that our model of emission dynamics and noise accurately describes the single photon emission and detection process. This enables us to further separate the relevant intrinsic photonic and technical sources of imperfection by means of the parameters extracted from the fit. We find pure dephasing of the emitter smaller than $1.0\,$MHz and spectral diffusion smaller than $3.8\,$MHz, consistent with the Fourier-limited linewidth of the emitter (Fig. \ref{fig1}), while the main remaining technical imperfection is the polarization mismatch of about 7° introduced by the MZI (see Supplementary Information). 
Considering only these three imperfections and the quantified background contributions we obtain a background-corrected visibility of $V_{\mathrm{HOM}}^{\mathrm{bc},93\%}=0.979_{-0.009}^{+0.006}$ ($V_{\mathrm{HOM}}^{\mathrm{bc},65\%}=0.980_{-0.009}^{+0.007}$).  Further correcting for the polarization mismatch and  the non zero intrinsic $g^{(2)}(0)$ value, we obtain $M^{\mathrm{intr},93\%}=0.997_{-0.011}^{+0.003}$ ($M^{\mathrm{intr},65\%}=0.999_{-0.012}^{+0.001}$) as the intrinsic photonic indistinguishability of the SnV center. A comprehensive list of all fitting parameters is given in Extended Data Table \ref{EDF:Fit_results}.

\section{Frequency-Conversion to the Telecom Band}
\begin{figure}[htbp]
  \centering
    \begin{subfigure}[t]{0.48\textwidth}
    \centering
    \includegraphics[width=0.98\textwidth]{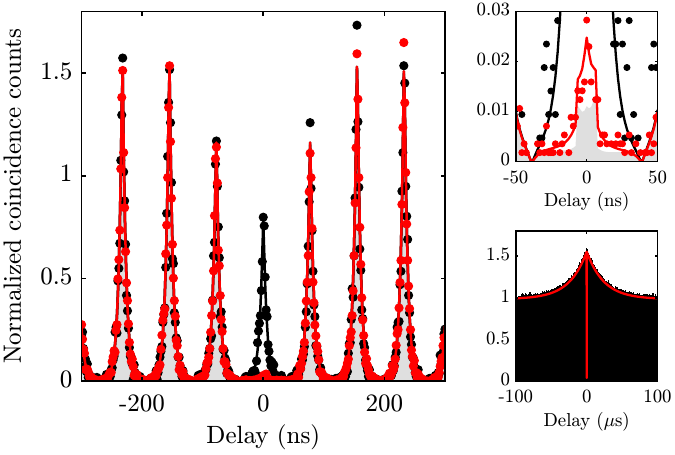}
    \caption{}
    \label{fig2:hom_vis77}
  \end{subfigure}
    \hfill
  \begin{subfigure}[t]{0.48\textwidth}
    \centering
    \includegraphics[width=0.98\textwidth]{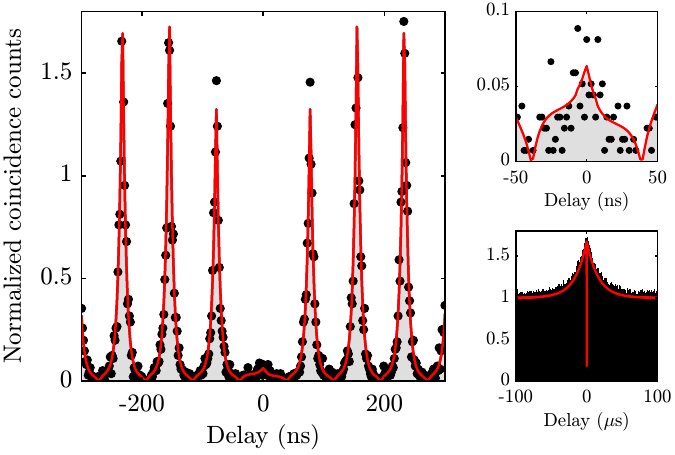}
    \caption{}
    \label{fig2:hom_tel77}
  \end{subfigure}
  \caption{\textbf{a}, Two-photon HOM interference of resonantly excited photons with repetition period $T_{\mathrm{rep}} = 77.5\,\mathrm{ns}$ and using the 93\% time-gating window. Consecutive photons are overlapped in an unbalanced fiber-based MZI with orthogonal (black) or parallel (red) polarization. Comparison of the central peak areas yields a raw visibility of \(V_\mathrm{HOM}^\mathrm{raw,93\%} = 0.950_{-0.008}^{+0.006}\). The grey shaded area represents the expected coincidence distribution for perfectly indistinguishable emitter photons, taking into account the measured background noise.
  \textbf{b}, Two-photon HOM interference after frequency conversion, resulting in a raw visibility of \(V_\mathrm{HOM}^\mathrm{raw,93\%} = 0.80_{-0.03}^{+0.01}\).}  
  \label{fig2}
\end{figure}
\begin{figure}[htbp]
  \centering
    \includegraphics[width=0.48\textwidth]{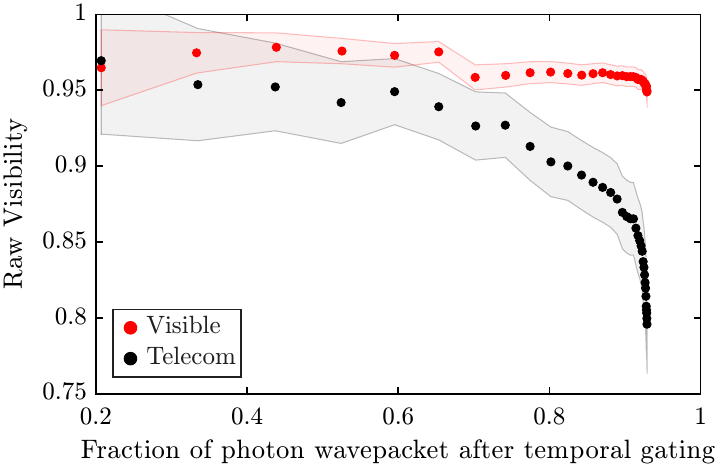}
  \caption{Raw two-photon interference visibility in the visible (red) and telecom (black) regime as a function of the remaining fraction of the photon wave packet after temporal gating of the detection window. The visibility is extracted from the raw coincidence counts by comparing the integrated area of the central peak to those of the far-delayed peaks. Shaded regions denote the 95\% confidence interval, which increases for narrower gating windows due to reduced photon statistics.}
  \label{fig3}
\end{figure}

Long-distance fiber networks require efficient and low-noise frequency conversion of visible photons to telecom wavelengths. We employ quantum frequency conversion of the SnV-resonant photons at 619\,nm to the telecom C-band at 1550\,nm in a two-stage conversion scheme \cite{Esfandyarpour18,Schaefer25}. Our device achieves a high fiber-to-fiber efficiency of 53\,\%, reduced to 43\,\% after narrowband spectral filtering, which suppresses conversion-induced noise to a very low rate of 22\,cps (see Methods).\\
To evaluate the indistinguishability of the converted photons, we again utilize an unbalanced MZI with a 77.5\,ns delay, following the same measurement procedure as for the visible regime. The correlation measurement for the time-gating window $t^\mathrm{G2}_{93\%}$, containing 93\,\% of the photon wave packet, is shown in Fig. \ref{fig2:hom_tel77}. The raw interference visibility is reduced to \(V_\mathrm{HOM}^\mathrm{raw,93\%} = 0.80_{-0.03}^{+0.01}\) compared to unconverted photons due to background contributions from converter noise and dark counts of telecom single-photon detectors of 15\,cps per channel. By applying a narrower temporal filtering, we observe a significant increase in SBR. Hence, gating with the second time window with a cutoff at $t^\mathrm{G2}_{65\%} = 7.3\,\mathrm{ns}$, results in a raw HOM visibility of \(V_\mathrm{HOM}^\mathrm{raw,65\%} = 0.940_{-0.029}^{+0.004}\), approaching the visibility observed for the unconverted photons.\\
Applying the same theoretical model as for the unconverted case with independently measured noise parameters gives a background-corrected visibility of $V_{\mathrm{HOM}}^{\mathrm{bc},93\%} =0.987_{-0.034}^{+0.009}$ ($V_{\mathrm{HOM}}^{\mathrm{bc},65\%} =0.994_{-0.031}^{+0.05}$) and an intrinsic indistinguishability of $M^{\mathrm{intr},93\%} = 0.992_{-0.031}^{+0.08} \quad \qty(M^{\mathrm{intr},65\%} =0.999_{-0.026}^{+0.001})$. All fitted parameters are listed in Extended Data Table \ref{EDF:Fit_results}. These values indicate that the reduction in raw visibility can be fully attributed to additional noise, while the intrinsic indistinguishability remains close to unity, demonstrating that the frequency converter does not degrade the indistinguishability beyond the reduction induced by the decreased SBR.\\
Fig. \ref{fig3} shows the raw HOM visibilities as a function of the applied temporal gating, expressed as the remaining fraction of the photon wave packet. This fraction is limited to values below 0.93, as the start of the detection window is chosen to exclude the residual excitation pulse. In the telecom regime, the visibility exhibits a pronounced dependence on the gating window, reflecting the elevated noise floor. By contrast, the measurement in the visible regime shows only a weak dependence on temporal filtering indicating a small uncorrelated background contribution. In both cases, a distinct increase in visibility is observed when the window is reduced to 0.65, corresponding to the complete exclusion of multi-path laser reflections from the detection window, highlighting their dominant contribution to the residual background.

\section{Discussion and Outlook}
\label{par_discus}
Our results establish SnV centers in diamond as a competitive platform for generating highly indistinguishable single photons suitable for quantum network applications. By combining resonant pulsed excitation, polarization-based filtering, and temporal gating, we achieve raw HOM visibilities exceeding 0.95. A central outcome of this work is the clear separation between intrinsic emitter properties and technical limitations. Our analysis shows that the measured two-photon interference visibility is not limited by intrinsic emitter-environment interactions, but by experimentally controllable imperfections such as residual excitation laser leakage, interferometer polarization mismatch, and detector background. After correcting for these technical contributions, we find negligible pure dephasing and spectral diffusion on the relevant timescale, consistent with the nearly Fourier-limited linewidth of the emitter, yielding intrinsic indistinguishabilities up to \(99.9\,\%\).\\
Importantly, these conclusions extend to the telecom wavelength regime via quantum frequency conversion. We find that the spectral and temporal properties of SnV photons are preserved throughout the conversion process, validating the suitability of this interface for fiber-based quantum links.\\
While we demonstrate indistinguishability using consecutively emitted photons from a single emitter, quantum-network operation requires interference between photons generated at remote and independent nodes. The negligible spectral diffusion and near-Fourier-limited emission observed here establish the key prerequisites for two-photon interference of independent sources. Moreover, distinct SnV centers have been tuned into mutual resonance \cite{Bushmakin2025,Waas2026}, and quantum frequency conversion provides an additional route for matching the photon frequencies of remote emitters \cite{Weber2019, Knaut2024}. Together, these results provide a clear path toward extending the intrinsic indistinguishability reported here to photons generated by independent SnV centers. \\
In quantum repeater architectures, photon indistinguishability limits the coherence of the heralded elementary links and hence the end-to-end performance. We quantify this with a Monte Carlo simulation of a memory-based repeater, built on the model of Ref. \cite{Kamin2023} and Barrett-Kok double-round heralded elementary-link generation \cite{Barrett2005,Bernien2013}, choosing typical values of 500 km link length and 30 segments. The simulation includes imperfect indistinguishability, background-induced false heralds, finite memory coherence and imperfect deterministic entanglement swapping, and evaluates the asymptotic BB84 secret-key rate; details are given in the Supplementary Information.
\begin{figure}[h]
   \centering
   \includegraphics[width=0.47\textwidth]{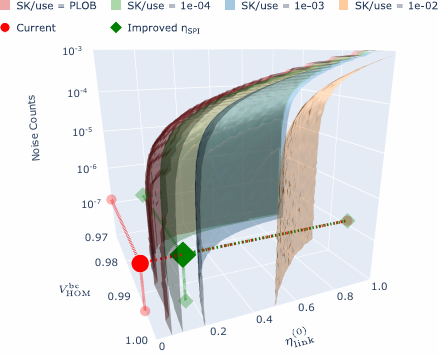}
   \caption{
   Monte Carlo simulation of repeater performance versus background-corrected HOM visibility \(V_{\rm HOM}^{\rm bc}\), single-photon detection efficiency  (excluding fiber losses) \(\eta_{\mathrm{link}}^{(0)}\) and total noise counts per detector and excitation round.  Iso-surfaces show parameter combinations resulting in the same secret key bits per channel use (SK/use). In the parallel protocol considered here, one repeater channel use corresponds to one simultaneous elementary-link attempt in all segments. Its duration \(\tau_{\rm cu}\) is limited by the signal travel time. 
The simulation assumes \(L=500\,{\rm km}\), \(n=30\) (therefore \(\tau_{\rm cu}=83.4\,\mu{\rm s}\)), \(T_{2,{\rm mem}}=1.35\,{\rm s}\) and \(F_{\rm swap}=0.999\). The red marker shows the measured \(65\,\%\) time-gated data; the green marker shows the projected performance for an improved collection efficiency corresponding to \(\eta_{\mathrm{link}}^{(0)}=0.172\).
The outermost iso-surface shows parameter combinations for which the SK/use equals the PLOB bound for direct transmission over \(500\,{\rm km}\) of \(1.94\times10^{-10}\).}
   \label{fig4}
\end{figure}
Figure \ref{fig4} shows the simulated performance compared to the Pirandola-Laurenza-Ottaviani-Banchi (PLOB) bound for direct transmission \cite{Pirandola2017} as a function of three experimentally accessible parameters: the background-corrected photon indistinguishability \(V_{\rm HOM}^{\rm bc}\), the single-photon detection efficiency \(\eta_{\mathrm{link}}^{(0)}\), which includes all efficiencies from photon generation to detection except fiber transmission, and the mean number of noise counts per detector and excitation round. We assume that spin-photon entanglement is generated on the electronic spin and immediately transferred to a nuclear-spin memory, for which a coherence time of \(T_{2,{\rm mem}}=1.35\,{\rm s}\) has been demonstrated in SnV centers \cite{Resch2026}. The swapping fidelity is set to \(F_{\rm swap}=0.999\), motivated by reported electron-nuclear spin gate fidelities of \(99.93\,\%\) \cite{Bartling2025} and readout fidelities of \(99.98\,\%\) \cite{Bhaskar2020} in color-center systems.\\
The indistinguishability and noise level demonstrated in our experiments already lie in the parameter regime required to exceed the repeaterless direct-transmission bound. It is important to note that surpassing the PLOB bound requires photon indistinguishabilities \(V_{\rm HOM}^{\rm bc}\gtrsim 97\,\%\), even for high efficiencies. In the current implementation, the dominant bottleneck is the collection efficiency of the emitted photons, which results in \(\eta_{\mathrm{link}}^{(0)} \approx 10^{-5}\). Importantly, this limitation is not intrinsic to the SnV center itself but arises from the absence of a cavity-enhanced spin–photon interface in the present experiment. Purcell-enhanced SnV nanocavities have been shown to channel 90\,\% \cite{Rugar2021} and 95\,\% \cite{Kuruma2021} of the excited-state decay into a zero-phonon-line cavity mode. Combining these values with the reported SnV quantum efficiency of 80\,\% \cite{Rugar2021}, a fiber-coupling efficiency of 90\,\% demonstrated for diamond nanophotonic devices \cite{Burek2017}, and the conversion, detection, and time-gating efficiencies achieved in this work suggests that a single-photon detection efficiency of $\eta_{\mathrm{link}}^{(0)}=17\,\%$ is feasible. With this improvement, the simulated repeater reaches $4.77\times10^{-4}$ secret bits per channel use, six orders of magnitude above the PLOB bound for a direct $500\,\mathrm{km}$ fiber link.\\
Notably, our analysis highlights the role of the optical interface, as we assume a near-unity spin–photon entanglement fidelity. While the development of high-fidelity spin–photon interfaces remains a central challenge, our simulation results emphasize that the fundamental performance requirements on photon indistinguishability, background level, and collection efficiency can be met, positioning SnV centers among the leading platforms for practical quantum repeater implementations.\\

\section*{Methods}
\subsection*{Sample Preparation}
\label{meth:sample} Two diamond samples, A and Iso1, were investigated. Both underwent identical ion implantation of $^{120}$Sn at 4\,MeV (implantation depth $\sim$2.5\,$\mu$m) with a dose of $10^{9}$\,ions\,cm$^{-2}$, followed by HPHT annealing at 2100$^\circ$C and 8\,GPa for 2\,h to mitigate implantation-induced damage. Sample A was an electronic-grade bulk diamond (Element Six) with a (100)-oriented polished surface, containing \textless5\,ppb substitutional nitrogen and \textless1\,ppb boron. Sample Iso1 was an electronic-grade diamond featuring a 50\,$\mu$m-thick isotopically enriched diamond layer. After HPHT treatment, Iso1 retained a rough surface morphology, whereas sample A was additionally subjected to chemical mechanical polishing (CMP) to remove surface roughness and produce a lower-defect surface. CMP was performed using a Logitech Tribo polishing system fitted with a SUBA-X polyester-impregnated polyurethane polishing pad. A commercially available colloidal silica polishing slurry (SF1) was used throughout the process, and the sample was polished for 10 hours. A detailed description of the polishing methodology is provided in \cite{Thomas2014}. A single SnV center in Sample Iso1 was used for all experiments, with the exception of the photoluminescence excitation measurement shown in Fig. \ref{fig1}, which was performed on Sample A.
\subsection*{Hong-Ou-Mandel Interferometry}
\label{meth:MZI}
To investigate the photonic indistinguishability of photons from a single SnV-color center, we repeatedly excite the emitter with an adjustable period $T_{\mathrm{rep}}$ and perform Hong-Ou-Mandel (HOM) measurements with the consecutively emitted fluorescence photons. This is achieved by sending
them through an unpolarized fiber Mach-Zehnder-Interferometer (MZI) with a path-difference
$\delta_{\mathrm{MZI}}$ (for technical realization see Supplementary Information) that compensates the temporal separation of the consecutive photon wave packets:
\begin{equation}
T_{\mathrm{rep}} =\delta_{\mathrm{MZI}}
\label{comp}
\end{equation}
Two-photon interference (TPI) occurs whenever two photons enter the second non-polarizing beamsplitter (BS) of the MZI through different input ports and is quantified via time-resolved coincidence measurements. To this end, superconducting nanowire single-photon detectors (SNSPDs) for visible (Single Quantum, timing jitter of 15\,ps and less than 1 dark counts/s) and telecom photons (ID Quantique, timing jitter less than 40\,ps and approximately 15 dark counts/s) are placed at the output ports of the HOM-BS and the time-tagged (Dotfast, timing jitter $<20$\,ps ) detection events are correlated to obtain time-resolved coincidence statistics. \\
Overlapping orthogonal, i.e. perfectly distinguishable photons, in the MZI leads to a 4:3:2:3:4 peak pattern, while for fully indistinguishable photons a 4:3:0:3:4 pattern with a vanishing central peak is expected due to HOM interference \cite{Loredo:16}.
Performing the experiment for both parallel ($\parallel$) and orthogonal ($\perp$) input polarization and extracting the central peak areas $A_{0,\parallel},A_{0,\perp}$ after renormalization with respect to the $4-$weights allows to evaluate the interference contrast 
\begin{equation}
    V_{\mathrm{HOM}} = 1 - \frac{A_{0,\parallel}}{A_{0,\perp}}=1 - \frac{p_{\parallel}}{p_{\perp}}~.
    \label{eq:VHOM}
\end{equation}
This reflects the ensemble-averaged wave packet overlap of the two photons interfering at the beamsplitter and, due to the shared normalization procedure, directly involves the overall probabilities $p_{\parallel},p_{\perp}$ of photon coincidences for parallel and orthogonal input polarizations, respectively.\\
When extracted directly from the raw data, this contrast is reduced by all imperfections inherent to both the single-photon source under investigation and the measurement apparatus, including statistical fluctuations of the single photon´s degrees of freedom as well as noise and background contributing to the coincidence statistics.\\
Understanding these reductions requires thorough quantitative modeling of the coincidence statistics by calculating the cross-correlation function $\mathcal{G}^{(2)}(\tau)$ obtained in the imperfect experiment (see next section). Since the various imperfections give rise to distinct temporal features in time-resolved correlation measurements, fitting $\mathcal{G}^{(2)}(\tau)$ to the data enables us to independently quantify the imperfections present in our experiment, while
the probabilities $p_{\parallel},p_{\perp}$ involved in Eq. \ref{eq:VHOM} are extracted by time-integration of the cross-correlation function. As described in the main text, the reference $p_{\perp}$ can either be obtained from the central peak of a second measurement with orthogonal polarization settings, or from the far delayed peaks of the same measurement with parallel-polarized photons. \\
In the following, we identify the dominant source-related and instrumental imperfections that directly affect the overlap of two single 
SnV-photons generated and interfered in our experiment. The role of background contributing to the coincidence statistics will be discussed in Methods section \ref{meth:theo}.\\ 
\\
\textit{Time--frequency degree of freedom:}\\
In accordance with Ref. \cite{Kambs2018}, we consider the emitter to be subject to environmental fluctuations that potentially affect its emission frequency statistically on two timescales: Perturbations with an interaction time longer than the repetition period $T_{\mathrm{rep}}$ of the experiment, giving rise to spectral diffusion (SD) and much faster fluctuations on a timescale of the photon lifetime and thereby leading to pure dephasing (PD) \cite{Kambs2018}. 
Thus, the single photons generated from such a perturbed single emitter form a statistical ensemble, with stochastic phase and frequency fluctuations leading to spectral broadening around a fixed carrier frequency.
In the time domain, fully overlapping modes of consecutive photons in each experimental cycle require both temporal stability and a vanishing relative delay, i.e. precise fulfillment of the condition given in Eq. \ref{comp}. Experimentally, the latter is well satisfied: By selectively opening one arm of the MZI, photons are routed exclusively through the remaining path. This procedure is performed for both configurations while monitoring photon arrival times via TCSPC. Fine adjustment of the repetition period $T_{\mathrm{rep}}$ with an arbitrary waveform generator (AWG-5064, Active Technologies), providing a timing resolution of below \SI{1}{\femto\second}, results in synchronous arrival, with a residual relative delay below the combined timing resolution of the SNSPDs and the time tagger.\\
With a constant relative delay effectively eliminated, the remaining temporal limitation arises from temporal stability. While the interferometer´s optical path-length difference can be affected by mechanical vibrations and thermal drifts, this can be neglected for the HOM-measurements for two reasons: 1) With fiber-length fluctuations on the \SI{}{\micro\meter} scale, the relative timing of photons arriving at the beamsplitter is affected only at the level of approximately $10^{-5}$ of the photon lifetime $\tau_{\mathrm{r}}$ and on timescales well above $T_{\mathrm{rep}}$. 2) Relative phase-differences of two photons impinging on the HOM-BS do not affect TPI, as they can be described as a global phase of the two-photon input state. Likewise, the coincidence statistics remain unaffected by relative phase differences owing to the near-complete population inversion achieved with our resonant $\pi$-pulse excitation, suppressing the effect of photon number coherence \cite{Loredo2019} (see Supplementary Information).\\
While for non-coherent, pulsed excitation schemes, the excitation process itself induces shot-to-shot variations in the relative excitation time of two single photons on the timescale of the pulse duration, the resonant excitation scheme presented in this work enables a coherent qubit control with near-unity $\pi$-pulse inversion efficiency, which does not suffer from a timing jitter. Consequently, we assume that the emitted photons share an identical temporal mode defined by the repeated drive–decay response, which, together with the fulfillment of Eq. \ref{comp}, ensures temporal indistinguishability of consecutively generated single SnV photons.\\
\\
\textit{Spatial degree of freedom:}\\
In the paraxial and narrowband regime, the longitudinal field structure is fully determined by the temporal (spectral) mode via the dispersion relation. Consequently, the spatial degree of freedom reduces to the transverse mode profile, which can be reliably overlapped for consecutive single photons by interfering them in an MZI implemented using identical single-mode fibers butt-coupled to evanescent 50:50 fiber couplers (Thorlabs).
This is evidenced by the high classical interference visibility of the home-built MZI, which is measured by periodically scanning the optical path difference between both MZI-arms with a piezo-electric fiber-stretcher (915-B, Evanescent Optics Inc.) in a range of several SnV-wavelengths.  
The resulting interference fringes are monitored at either output port of the MZI and statistical analysis of the fitted extrema yields a mean classical interference contrast of $V_{\mathrm{class.}}=99.2_{-1.2}^{+1.2}$ (for details see Supplementary Information).
This is fully explained by the finite first-order coherence of $g^{(1)}(\SI{77}{\nano\second})\approx 99\%$ resulting from the \SI{620}{nm} continuous wave probing laser (TA-SHG pro, Toptica, spectral width $\Delta \nu_{\mathrm{FWHM}}\approx \SI{40}{\kilo\hertz}$.), indicating that  imbalanced MZI-arm transmission, as well as any deviation of the beamsplitter reflectivities and transmissivities from $\mathcal{R}=\mathcal{T}$ are negligible (For details see Supplementary Information).\\
\\
\textit{Polarization degree of freedom:}\\
Despite the dipole C-transition exhibiting a well-defined linear emission polarization \cite{Gorlitz2020}, single photons propagating along different MZI paths may undergo distinct polarization transformations due to birefringence in the optical fibers. In the present work, standard single-mode fibers were used, with one set of home-built, paddle-based polarization controllers allowing to adjust the global input polarization  and two more defining the polarization states independently for the two MZI-arms. This provided additional degrees of freedom in optimizing the MZI for both configurations of parallel and orthogonal polarization input states at the HOM-BS (see Supplementary Information). While this enables the high classical visibility demonstrated above, thermal drifts gradually introduce a slight polarization mismatch during multi-hour correlation measurements. Fixing the spatial alignment of all MZI components, once adjusted, and shielding the interferometer with a styrofoam enclosure suppresses this polarization drift to an effective value of $\beta_{\mathrm{eff.}}\approx 8^{\circ}$, independently assessed from the time-averaged classical interferometer visibility (see Supplementary Information). This is in good agreement with the polarization mismatch $\beta=7^{\circ}$ presented in the main text, which is extracted by fitting our model for $\mathcal{G}^{(2)}(\tau)$ to the HOM-data. Polarization-maintaining fibers are expected to overcome this purely technical limitation, further enhancing the near-unity spatiotemporal mode overlap of consecutive SnV photons.
\subsection*{Theoretical modeling of joint detection probabilities for time-resolved photon correlation and HOM measurements}
\label{meth:theo}
To obtain analytical expressions modeling time resolved correlation statistics we follow the formalism presented in \cite{Legero_2003,Kambs2018}. 
Incorporating intrinsic photonic imperfections of the fluorescence photons by modeling the single photon wavefunction and extracting background contributions from TCSPC measurements we calculate the probability for a joint photon detection behind the beamsplitter at times $t_0$ and $t_0+\tau$, given the modeled input as well as the temporal gating presented in the main text.\\
As a starting point, we consider the background-free, ungated case discussed in Ref. \cite{Kambs2018} where two single photons impinge on a beamsplitter via different input ports \enquote{1} and \enquote{2}. With a beamsplitter reflectivity $\mathcal{R}$ and transmissivity $\mathcal{T}$, as well as overall detection efficiencies $\eta_3,\eta_4$ at the detectors placed behind the respective output ports \enquote{3} and \enquote{4} (see Supplementary Information), this yields \cite{Kambs2018}: 
\begin{align}
P_{\mathrm{joint}}(t_0,\!\tau)=&\eta_3\eta_4| \mathcal{T}\zeta_2(t_0)\zeta_1(t_0\!+\!\tau)\nonumber \\
+&\mathcal{R}\zeta_2(t_0\!+\!\tau)\zeta_1(t_0)|^2~,
\label{pjoint}
\end{align}
directly depending on the normalized field modes $\zeta_i(t)$ of the two photons
interacting at the beamsplitter. In agreement with Ref. \cite{Kambs2018}, we incorporate the stochastic effects of SD and PD (see Methods \ref{meth:MZI}) considering that the single-photon field modes $\zeta_i(t)$ exhibit statistically distributed frequency and phase. With this, the PD- and SD- averaged cross-correlation function
\begin{equation}
\mathcal{G}^{(2)}(\tau)\!=\!\int \! \mathrm{d}t_0\bigg\langle\! \bigg\langle 
\!\frac{P_{\mathrm{joint}}(t_0,\tau)}{\eta_3 \eta_4} \!\bigg\rangle\! \bigg\rangle_{\mathrm{PD,SD}}
\label{crosscorr}
\end{equation}
is calculated, being the observable assessed in time-resolved HOM-experiments.\\
As can be seen from the TCSPC measurement presented in Fig. \ref{fig1:TCSPC} of the main text, the realistic experiment results in a statistical ensemble of photons that contains background- and noise  contributions with distinct temporal features. Since these contributions also enter the cross-correlation of detection events in $g^{(2)}$ and HOM-measurements, they must be taken into account. Considering the dominant terms listed below, we model the raw input prior to the MZI as a statistical mixture, corresponding to a time-resolved photon detection probability comprising different components:
\begin{align}
p(t)^{\mathrm{\scriptsize raw}}_{\mathrm{\scriptsize gated}}
= &~W(t)\bigg[~\alpha ~\big(\underbrace{\alpha^{\mathrm{ s}} p^{\mathrm{s}}(t)+\alpha^{\mathrm{ cwe}}p^{\mathrm{cwe}}(t)}_{=~p^{\mathrm{SnV}}(t)}\big)\nonumber \\ 
&~~+ \sum_{k} \beta_k ~ p^\mathrm{ gauss}_{k}(t) +\gamma ~ p^{\mathrm{cw}}\bigg]~.
\label{I_mixture}
\end{align}
Directly detected in a TCSPC measurements with a gating window described by $W(t)$ (see below), this results in a counting statistics which is scaled by a global factor $\mathcal{N}$. We note that in Eq. \ref{I_mixture}, the relative component weights satisfy the normalization conditions\\
$\alpha + \sum_{k} \beta_k +\gamma = 1$ and $\alpha^{\mathrm{ s}}+\alpha^{\mathrm{ cwe}}=1$, 
while the individual components, explained in the following, are normalized according to:
\begin{equation}
    \int \mathrm{d}t~W(t)~p^i(t) =1~,
    \label{norm_pi}
\end{equation}
such that
\begin{align}
\int \mathrm{d}\mathrm{t} ~p(t)^{\mathrm{\scriptsize raw}}_{\mathrm{gated}} =1~.
\label{norm_p}
\end{align}
Consistent with the formalism discussed in Ref. \cite{Kambs2018}, the signal part $p^{\mathrm{s}}(t)$ of the SnV fluorescence, $p^{\mathrm{SnV}}(t)$, is obtained from modeling the single photons with the one-sided exponential $\zeta^{\mathrm{SnV}}(t)$ and accounting for the photodynamics of the pulsed excitation (for detailed expressions see Supplementary Information). As can be seen in Fig. \ref{fig1:TCSPC} in the main text, the dominant background contribution originates from multiple residual laser reflections, which are described by a sum of Gaussian functions $p^\mathrm{ gauss}_{k}(t)$. The finite extinction ratio of the EOM pulses gives rise to two contributions: First, between consecutive pulses, there remains a constant probability of detecting scattered laser photons. Combined with the constant background originating from detector dark counts, this contribution is described by a constant offset $p^{\mathrm{cw}}$. Second, the imperfectly suppressed laser gives rise to residual coherent cw-excitation with subsequent photon emission following the probability distribution $p_{\mathrm{e}}^{\mathrm{cwe}}(t)$ presented in Supplementary Information. We note, that the relative weights $\alpha^{\mathrm{ s}},\alpha^{\mathrm{ cwe}}$ of SnV fluorescence photons being emitted and detected due to either one of the two excitation processes are entirely fixed by the excitation pulse duration and the extinction ratio achieved with the EOM (see Supplementary Information).\\
Performing our purely post selective temporal gating in TCSPC measurements means that in each experimental cycle only detection events with times $t$ within the interval $0\leq t^{\mathrm{G1}}\leq t \leq t^{\mathrm{G2}}\leq T_{\mathrm{rep}} $ are retained. This is accounted for by the Heaviside-window $\mathrm{W}(t):=\mathrm{H} (t-t^{\mathrm{G1}})\mathrm{H} (t^{\mathrm{G2}}-t)$ in Eq. \ref{I_mixture} and the renormalization according to Eq. \ref{norm_pi}
The resulting probability distribution is fitted to the TCSPC data with the relative component weights $\alpha, \beta_{k},\gamma$, as well as the global scaling factor $\mathcal{N}$ serving as free fit parameters. This enables direct extraction of the signal-to-background ratios prior to the MZI, which subsequently enter as fixed input parameters when modeling the $g^{(2)}$- and HOM-measurements (for details see Supplementary Information).\\
Finally, the temporal gating is consistently incorporated throughout the correlation model by applying appropriate Heaviside windows and renormalization to all single-photon field modes and background contributions entering Eq. \ref{pjoint}, such that evaluating Eq. \ref{crosscorr} yields the gated cross-correlation function  $\mathcal{G}^{(2)}_{\mathrm{gated}}(\tau)$.
This preserves the model´s ability of independently quantifying emitter-related and technical imperfections, whose distinct temporal signatures in the HOM-data are further modulated when the cross-correlation statistics is obtained from the gated detector signals.\\
Explicit analytical expressions can be found in Supplementary Information. Details on the derivation will be published elsewhere.
\subsection*{Quantum Frequency Conversion}
\label{meth:QFC}
 Frequency conversion is performed via difference-frequency generation in 40‑mm-long PPLN waveguides in a two-stage scheme \cite{Esfandyarpour18,Schaefer25}, that suppresses conversion-induced noise at the target wavelength, following the experimental design employed in Ref. \cite{Schaefer25}. By mixing with a strong 2062\,nm pump beam, photons at 619\,nm are converted to the telecom C band (1550\,nm) via an intermediate wavelength of 885\,nm. As pump laser we employ a  GaSb-based vertical-external-cavity surface-emitting laser (VECSEL) built by Fraunhofer IAF \cite{Holl2025}, actively stabilized to an optical frequency comb (Menlo Systems). The quantum frequency converter reaches internal conversion efficiencies of 88\,\% (90\,\%) for the first (second) conversion stage, measured via signal depletion of a coherent laser at low power. Including coupling and reflection losses, we obtain a fiber-to-fiber external efficiency of 53\,\%, which is reduced to 43\,\% upon addition of a 9.5\,GHz FWHM narrowband spectral filter set-up composed of two cascaded volume Bragg gratings. Maximum conversion efficiency requires an input pump power of 680\,mW, inducing a noise rate of 22\,cps within the filter bandwidth. Correcting for the detection efficiency of the SNSPDs and optical losses in the detection path yields a device noise count rate of 45\,photons/s (4.7\,photons/s per GHz filter bandwidth). Further narrowing of the filter bandwidth would induce temporal broadening of the residual excitation pulse leaking through the cross-polarization filtering, thereby increasing its temporal overlap with the emitted photon wave packet.

\section*{Acknowledgement}
This research received funding from the German Federal Ministry of Research, Technology and Space (BMFTR) within the projects QR.X (Contract No. 16KISQ001K), QR.N (Contract No. 16KIS2180K), QPIC-1 (Contract No. 13N15859) and HiFi (Contract No. 13N15926); and the Deutsche Forschungsgemeinschaft (DFG, German Research Foundation) – Project-ID 429529648 – TRR 306 QuCoLiMa (“Quantum Cooperativity of Light and Matter”). 
We thank Detleff Rogalla (Rubion) for ion implantation of the samples. We thank Peter Michler, Richard Warburton, Klaus Jöns, Alexander Högele, Ilja Gerhard, Florian Kaiser, Simone Portalupi, Hubert Lam, Jens-Peter Ruske and Thomas Kinder for helpful discussions.
We thank Johannes Görlitz and Benjamin Kambs for contributions to the early stage of the experiment.

\section*{Author Contributions}

D.H. and R.M.-G. conceived the experiment, built the setup for resonant excitation of single SnV centers and the MZI for temporal overlap of consecutive photons, and performed the measurements on the non-frequency-converted photons. D.H. and L.E. performed the long-term PLE as well as the Rabi oscillation measurements of the emitter. T.B., M.S. and D.L. conceived the quantum frequency conversion setup, which was built by M.S. and D.L. D.H. and T.B. built the telecom MZI and performed the telecom two-photon interference measurements together with M.S. and D.L., while D.H. and T.B. evaluated the data with input from R.M.-G., M.S., D.L. and C.B.. R.M.-G., D.H. and C.B. developed and refined the theoretical model describing the two-photon interference. D.H. and L.E. implemented the MC simulations. T.B., D.H., R.M.-G., M.S., D.L. and P.v.L. developed and discussed the repeater link simulations. M.M. and N.P. prepared the diamond sample and carried out the HPHT annealing. S.M. and O.W. performed the surface polishing of the sample. C.B. supervised the project. The manuscript was written by D.H., R.M.-G., T.B., M.S. and D.L., with input from all authors.

\section*{Data availability}

\section*{Competing interests}
T.B., M.S., D.L. and C.B. are involved in developing quantum frequency conversion technology at OPTIQAL Quantum Technologies. The other authors declare no competing interests.


\bibliography{references}

@article{Gorlitz2020,
doi = {10.1088/1367-2630/ab6631},
year = {2020},
month = {jan},
publisher = {IOP Publishing},
volume = {22},
number = {1},
pages = {013048},
author = {Görlitz, Johannes and Herrmann, Dennis and Thiering, Gergő and Fuchs, Philipp and Gandil, Morgane and Iwasaki, Takayuki and Taniguchi, Takashi and Kieschnick, Michael and Meijer, Jan and Hatano, Mutsuko and Gali, Adam and Becher, Christoph},
title = {Spectroscopic investigations of negatively charged tin-vacancy centres in diamond},
journal = {New J. Phys.},
}

@article{Gorlitz2022,
  title = {Coherence of a charge stabilised tin-vacancy spin in diamond},
  author = {G\"orlitz, J. and Herrmann, D. and Fuchs, P. and Miyazaki, T. and Kato, H. and Mizuochi, N. and Isoya, J. and Iwasaki, T. and Jelezko, F. and Siyushev, P.},
  journal = {npj Quantum Inf.},
  volume = {8},
  number = {1},
  pages = {45},
  year = {2022},
  doi = {10.1038/s41534-022-00552-0}
}

@ARTICLE{Fischer2016,
  title     = "Dynamical modeling of pulsed two-photon interference",
  author    = "Fischer, Kevin A and M{\"u}ller, Kai and Lagoudakis,
               Konstantinos G and Vu{\v c}kovi{\'c}, Jelena",
  journal   = "New J. Phys.",
  publisher = "IOP Publishing",
  volume    =  18,
  number    =  11,
  pages     = "113053",
  month     =  nov,
  year      =  2016,
  copyright = "http://creativecommons.org/licenses/by/3.0/"
}

@article{Hong1987,
  title = {Measurement of subpicosecond time intervals between two photons by interference},
  author = {Hong, C. K. and Ou, Z. Y. and Mandel, L.},
  journal = {Phys. Rev. Lett.},
  volume = {59},
  number = {18},
  pages = {2044--2046},
  year = {1987},
  doi = {10.1103/PhysRevLett.59.2044}
}

@article{Esfandyarpour18,
author = {Vahid Esfandyarpour and Carsten Langrock and Martin Fejer},
journal = {Opt. Lett.},
number = {22},
pages = {5655--5658},
publisher = {Optica Publishing Group},
title = {Cascaded downconversion interface to convert single-photon-level signals at 650 nm to the telecom band},
volume = {43},
month = {Nov},
year = {2018},
doi = {10.1364/OL.43.005655},
}

@article{Kuhlmann2013,
    author = {Kuhlmann, Andreas V. and Houel, Julien and Brunner, Daniel and Ludwig, Arne and Reuter, Dirk and Wieck, Andreas D. and Warburton, Richard J.},
    title = {A dark-field microscope for background-free detection of resonance fluorescence from single semiconductor quantum dots operating in a set-and-forget mode},
    journal = {Rev. Sci. Instrum.},
    volume = {84},
    number = {7},
    pages = {073905},
    year = {2013},
    month = {07},
    issn = {0034-6748},
    doi = {10.1063/1.4813879},
}

@article{Schaefer25,
author = {Schäfer, Marlon and Kambs, Benjamin and Herrmann, Dennis and Bauer, Tobias and Becher, Christoph},
title = {Two-Stage, Low Noise Quantum Frequency Conversion of Single Photons from Silicon-Vacancy Centers in Diamond to the Telecom {C}-Band},
journal = {Adv. Quantum Technol.},
volume = {8},
number = {2},
pages = {2300228},
doi = {https://doi.org/10.1002/qute.202300228},
year = {2025}
}

@article{Somaschi2016,
   title={Near-optimal single-photon sources in the solid state},
   volume={10},
   ISSN={1749-4893},
   DOI={10.1038/nphoton.2016.23},
   number={5},
   journal={Nat. Photon.},
   publisher={Springer Science and Business Media LLC},
   author={Somaschi, N. and Giesz, V. and De Santis, L. and Loredo, J. C. and Almeida, M. P. and Hornecker, G. and Portalupi, S. L. and Grange, T. and Antón, C. and Demory, J. and Gómez, C. and Sagnes, I. and Lanzillotti-Kimura, N. D. and Lemaítre, A. and Auffeves, A. and White, A. G. and Lanco, L. and Senellart, P.},
   year={2016},
   month=mar, pages={340–345} 
}

@article{Wehner2018,
  author  = {Wehner, Stephanie and Elkouss, David and Hanson, Ronald},
  title   = {Quantum internet: A vision for the road ahead},
  journal = {Science},
  volume  = {362},
  number  = {6412},
  pages   = {9288},
  year    = {2018},
  doi     = {10.1126/science.aam9288}
}

@article{Briegel1998,
  author  = {Briegel, H.-J. and D{\"u}r, W. and Cirac, J. I. and Zoller, P.},
  title   = {Quantum Repeaters: The Role of Imperfect Local Operations in Quantum Communication},
  journal = {Phys. Rev. Lett.},
  volume  = {81},
  number  = {26},
  pages   = {5932--5935},
  year    = {1998},
  doi     = {10.1103/PhysRevLett.81.5932}
}

@article{Pan1998,
  author  = {Pan, J.-W. and Bouwmeester, D. and Weinfurter, H. and Zeilinger, A.},
  title   = {Experimental Entanglement Swapping: Entangling Photons That Never Interacted},
  journal = {Phys. Rev. Lett.},
  volume  = {80},
  number  = {18},
  pages   = {3891--3894},
  year    = {1998},
  doi     = {10.1103/PhysRevLett.80.3891}
}

@article{Cabrillo1999,
  author  = {Cabrillo, C. and Cirac, J. I. and Garcia-Fernandez, P. and Zoller, P.},
  title   = {Creation of entangled states of distant atoms by interference},
  journal = {Phys. Rev. A},
  volume  = {59},
  pages   = {1025},
  year    = {1999},
  doi     = {10.1103/PhysRevA.59.1025}
}

@article{Simon2003,
  author  = {Simon, Christoph and Irvine, William T. M.},
  title   = {Robust Long-Distance Entanglement and a Loophole-Free Bell Test with Ions and Photons},
  journal = {Phys. Rev. Lett.},
  volume  = {91},
  number  = {11},
  pages   = {110405},
  year    = {2003},
  doi     = {10.1103/PhysRevLett.91.110405}
}

@article{Bhaskar2020,
	title = {Experimental demonstration of memory-enhanced quantum communication},
	volume = {580},
	issn = {1476-4687},
	doi = {10.1038/s41586-020-2103-5},
	language = {en},
	number = {7801},
	journal = {Nature},
	publisher = {Nature Publishing Group},
	author = {Bhaskar, M. K. and Riedinger, R. and Machielse, B. and Levonian, D. S. and Nguyen, C. T. and Knall, E. N. and Park, H. and Englund, D. and Lončar, M. and Sukachev, D. D. and Lukin, M. D.},
	year = {2020},
	pages = {60--64},
}

@article{Knaut2024,
	title = {Entanglement of nanophotonic quantum memory nodes in a telecom network},
	volume = {629},
	copyright = {2024 The Author(s)},
	issn = {1476-4687},
	doi = {10.1038/s41586-024-07252-z},
	number = {8012},
	urldate = {2026-05-26},
	journal = {Nature},
	publisher = {Nature Publishing Group},
	author = {Knaut, C. M. and Suleymanzade, A. and Wei, Y.-C. and Assumpcao, D. R. and Stas, P.-J. and Huan, Y. Q. and Machielse, B. and Knall, E. N. and Sutula, M. and Baranes, G. and Sinclair, N. and De-Eknamkul, C. and Levonian, D. S. and Bhaskar, M. K. and Park, H. and Lončar, M. and Lukin, M. D.},
	year = {2024},
	pages = {573--578},

}

@article{Ruf2021,
  author  = {Ruf, Maximilian and Wan, Noel H. and Choi, Hyeongrak and Englund, Dirk and Hanson, Ronald},
  title   = {Quantum networks based on color centers in diamond},
  journal = {J. Appl. Phys.},
  volume  = {130},
  number  = {7},
  pages   = {070901},
  year    = {2021},
  doi     = {10.1063/5.0056534}
}

@article{Wei2022,
  author  = {Wei, Shi-Hai and Jing, Bo and Zhang, Xue-Ying and Liao, Jin-Yu and Yuan, Chen-Zhi and Fan, Bo-Yu and Lyu, Chen and Zhou, Dian-Li and Wang, You and Deng, Guang-Wei and Song, Hai-Zhi and Oblak, Daniel and Guo, Guang-Can and Zhou, Qiang},
  title   = {Towards Real-World Quantum Networks: A Review},
  journal = {Laser Photon. Rev.},
  volume  = {16},
  number  = {3},
  pages   = {2100219},
  year    = {2022},
  doi     = {10.1002/lpor.202100219}
}

@article{Kiraz2004,
  author  = {Kiraz, Alper and Atat{\"u}re, Mete and Imamo{\u{g}}lu, Atac},
  title   = {Quantum-dot single-photon sources: Prospects for applications in linear optics quantum-information processing},
  journal = {Phys. Rev. A},
  volume  = {69},
  pages   = {032305},
  year    = {2004},
  doi     = {10.1103/PhysRevA.69.032305}
}

@article{Kambs2018,
  author  = {Kambs, Benedikt and Becher, Christoph},
  title   = {Limitations on the indistinguishability of photons from solid-state single-photon sources},
  journal = {New J. Phys.},
  volume  = {20},
  pages   = {115003},
  year    = {2018},
  doi     = {10.1088/1367-2630/aaea99}
}

@article{Meraner2020,
  title = {Indistinguishable photons from a trapped-ion quantum network node},
  author = {Meraner, M. and Mazloom, A. and Krutyanskiy, V. and Krcmarsky, V. and Schupp, J. and Fioretto, D. A. and Sekatski, P. and Northup, T. E. and Sangouard, N. and Lanyon, B. P.},
  journal = {Phys. Rev. A},
  volume = {102},
  issue = {5},
  pages = {052614},
  numpages = {14},
  year = {2020},
  month = {Nov},
  publisher = {American Physical Society},
  doi = {10.1103/PhysRevA.102.052614},
}

@article{Bernien2013,
  title = {Heralded entanglement between solid-state qubits separated by three metres},
  author = {Bernien, H. and Hensen, B. and Pfaff, W. and Koolstra, G. and Blok, M. S. and Robledo, L. and Taminiau, T. H. and Markham, M. and Twitchen, D. J. and Childress, L. and Hanson, R.},
  journal = {Nature},
  volume = {497},
  number = {7447},
  pages = {86--90},
  year = {2013},
  doi = {10.1038/nature12016}
}

@article{Moehring2007,
  author  = {Moehring, D. L. and others},
  title   = {Entanglement of single-atom quantum bits at a distance},
  journal = {Nature},
  volume  = {449},
  pages   = {68--71},
  year    = {2007},
  doi     = {10.1038/nature06118}
}

@article{Hofmann2012,
  author  = {Hofmann, J. and others},
  title   = {Heralded entanglement between widely separated atoms},
  journal = {Science},
  volume  = {337},
  pages   = {72--75},
  year    = {2012},
  doi     = {10.1126/science.1221856}
}

@article{Delteil2016,
  author  = {Delteil, Aymeric and others},
  title   = {Generation of heralded entanglement between distant hole spins},
  journal = {Nat. Phys.},
  volume  = {12},
  pages   = {218--223},
  year    = {2016},
  doi     = {10.1038/nphys3605}
}

@article{Munro2012,
  author  = {Munro, W. J. and Stephens, A. M. and Devitt, S. J. and Harrison, K. A. and Nemoto, Kae},
  title   = {Quantum communication without the necessity of quantum memories},
  journal = {Nat. Photon.},
  volume  = {6},
  number  = {11},
  pages   = {777--781},
  year    = {2012},
  doi     = {10.1038/nphoton.2012.243}
}

@article{Azuma2015,
  author  = {Azuma, Koji and Tamaki, Kiyoshi and Lo, Hoi-Kwong},
  title   = {All-photonic quantum repeaters},
  journal = {Nat. Commun.},
  volume  = {6},
  pages   = {6787},
  year    = {2015},
  doi     = {10.1038/ncomms7787}
}

@article{Lindner2009,
  author  = {Lindner, Netanel H. and Rudolph, Terry},
  title   = {Proposal for Pulsed On-Demand Sources of Photonic Cluster State Strings},
  journal = {Phys. Rev. Lett.},
  volume  = {103},
  pages   = {113602},
  year    = {2009},
  doi     = {10.1103/PhysRevLett.103.113602}
}

@article{Thomas2022,
  author  = {Thomas, Philip and Ruscio, Leonardo and Morin, Olivier and Rempe, Gerhard},
  title   = {Efficient generation of entangled multi-photon graph states from a single atom},
  journal = {Nature},
  volume  = {608},
  pages   = {677--681},
  year    = {2022},
  doi     = {10.1038/s41586-022-04987-5}
}

@article{Cogan2023,
  author  = {Cogan, Dan and Su, Zu-En and Kenneth, Oded and Gershoni, David},
  title   = {Deterministic generation of indistinguishable photons in a cluster state},
  journal = {Nat. Photon.},
  volume  = {17},
  number  = {4},
  pages   = {324--329},
  year    = {2023},
  doi     = {10.1038/s41566-022-01152-2}
}

@article{Economou2010,
  author  = {Economou, Sophia E. and Lindner, Netanel H.},
  title   = {Optically Generated 2-Dimensional Photonic Cluster State from Coupled Quantum Dots},
  journal = {Phys. Rev. Lett.},
  volume  = {105},
  pages   = {093601},
  year    = {2010},
  doi     = {10.1103/PhysRevLett.105.093601}
}

@article{Thomas2024,
  author  = {Thomas, Philip and Ruscio, Leonardo and Morin, Olivier and Rempe, Gerhard},
  title   = {Fusion of deterministically generated photonic graph states},
  journal = {Nature},
  volume  = {629},
  pages   = {567--572},
  year    = {2024},
  doi     = {10.1038/s41586-024-07357-5}
}

@article{Istrati2020,
	title = {Sequential generation of linear cluster states from a single photon emitter},
	volume = {11},
	copyright = {2020 The Author(s)},
	issn = {2041-1723},
	doi = {10.1038/s41467-020-19341-4},
	language = {en},
	number = {1},
	journal = {Nat. Commun.},
	publisher = {Nature Publishing Group},
	author = {Istrati, D. and Pilnyak, Y. and Loredo, J. C. and Antón, C. and Somaschi, N. and Hilaire, P. and Ollivier, H. and Esmann, M. and Cohen, L. and Vidro, L. and Millet, C. and Lemaître, A. and Sagnes, I. and Harouri, A. and Lanco, L. and Senellart, P. and Eisenberg, H. S.},
	month = oct,
	year = {2020},
	pages = {5501},
}

@article{Bartolucci2023,
  author  = {Bartolucci, Sara and Birchall, Patrick and Bomb{\'i}n, Hector and Cable, Hugo and Dawson, Chris and Gimeno-Segovia, Mercedes and Johnston, Eric and Kieling, Konrad and Nickerson, Naomi and Pant, Mihir and Pastawski, Fernando and Rudolph, Terry and Sparrow, Chris},
  title   = {Fusion-based quantum computation},
  journal = {Nat. Commun.},
  volume  = {14},
  number  = {1},
  pages   = {912},
  year    = {2023},
  doi     = {10.1038/s41467-023-36493-1}
}

@article{Kamin2023,
  author  = {Kamin, L. and Shchukin, E. and Schmidt, F. and van Loock, P.},
  title   = {Exact rate analysis for quantum repeaters with imperfect memories and entanglement swapping as soon as possible},
  journal = {Phys. Rev. Res},
  volume  = {5},
  number  = {2},
  pages   = {023086},
  year    = {2023},
  doi     = {10.1103/PhysRevResearch.5.023086}
}

@article{Chan2025,
  author  = {Chan, Ming Lai and Bell, Thomas J. and Pettersson, Love A. and Chen, Susan X. and Yard, Patrick and S{\o}rensen, Anders S. and Paesani, Stefano},
  title   = {Tailoring Fusion-Based Photonic Quantum Computing Schemes to Quantum Emitters},
  journal = {PRX Quantum},
  volume  = {6},
  pages   = {020304},
  year    = {2025},
  doi     = {10.1103/PRXQuantum.6.020304},
  eprint  = {2410.06784},
  archivePrefix = {arXiv},
  primaryClass  = {quant-ph}
}

@article{Couteau2023,
  author  = {Couteau, Christophe and Barz, Stefanie and Durt, Thomas and Gerrits, Thomas and others},
  title   = {Applications of single photons to quantum communication and computing},
  journal = {Nat. Rev. Phys.},
  volume  = {5},
  number  = {6},
  pages   = {326--338},
  year    = {2023},
  doi     = {10.1038/s42254-023-00583-2}
}

@article{Giorgino2025,
    author = {F. Giorgino and P. Zah\'{a}lka and L. Jehle and L. Carosini and L. M. Hansen and J. C. Loredo and P. Walther},
    journal = {Opt. Quantum},
    number = {4},
    pages = {402--407},
    publisher = {Optica Publishing Group},
    title = {Multi-photon emission from a resonantly pumped quantum dot},
    volume = {3},
    month = {Aug},
    year = {2025},
    doi = {10.1364/OPTICAQ.557604},
}

@article{Debroux2021,
  author  = {Debroux, Romain and others},
  title   = {Quantum Control of the Tin-Vacancy Spin Qubit in Diamond},
  journal = {Phys. Rev. X},
  volume  = {11},
  pages   = {041041},
  year    = {2021},
  doi     = {10.1103/PhysRevX.11.041041}
}

@article{Trusheim2020,
  author  = {Trusheim, Matthew E. and others},
  title   = {Transform-Limited Photons From a Coherent Tin-Vacancy Spin in Diamond},
  journal = {Phys. Rev. Lett.},
  volume  = {124},
  pages   = {023602},
  year    = {2020},
  doi     = {10.1103/PhysRevLett.124.023602}
}

@article{Karapatzakis2024,
  title = {Microwave Control of the Tin-Vacancy Spin Qubit in Diamond with a Superconducting Waveguide},
  author = {Karapatzakis, Ioannis and Resch, Jeremias and Schrodin, Marcel and Fuchs, Philipp and Kieschnick, Michael and Heupel, Julia and Kussi, Luis and S\"urgers, Christoph and Popov, Cyril and Meijer, Jan and Becher, Christoph and Wernsdorfer, Wolfgang and Hunger, David},
  journal = {Phys. Rev. X},
  volume = {14},
  issue = {3},
  pages = {031036},
  numpages = {24},
  year = {2024},
  month = {Aug},
  publisher = {American Physical Society},
  doi = {10.1103/PhysRevX.14.031036},
}

@article{Resch2026,
  title = {High-Fidelity Control of a $^{13}\mathrm{C}$ Nuclear Spin Coupled to a Tin-Vacancy Center in Diamond},
  author = {Resch, Jeremias and Karapatzakis, Ioannis and Elshorbagy, Mohamed and Schrodin, Marcel and Fuchs, Philipp and Gra\ss{}hoff, Philipp and Kussi, Luis and S\"urgers, Christoph and Popov, Cyril and Becher, Christoph and Wernsdorfer, Wolfgang and Hunger, David},
  journal = {Phys. Rev. X},
  volume = {16},
  issue = {1},
  pages = {011060},
  numpages = {18},
  year = {2026},
  month = {Mar},
  publisher = {American Physical Society},
  doi = {10.1103/bmc6-qvwq},
}

@article{tomm2021,
	title = {A bright and fast source of coherent single photons},
	volume = {16},
	issn = {1748-3387, 1748-3395},
	doi = {10.1038/s41565-020-00831-x},
	language = {en},
	number = {4},
	urldate = {2026-04-24},
	journal = {Nat. Nanotechnol.},
	author = {Tomm, Natasha and Javadi, Alisa and Antoniadis, Nadia Olympia and Najer, Daniel and Löbl, Matthias Christian and Korsch, Alexander Rolf and Schott, Rüdiger and Valentin, Sascha René and Wieck, Andreas Dirk and Ludwig, Arne and Warburton, Richard John},
	month = apr,
	year = {2021},
	pages = {399--403},
}

@article{van_leent_entangling_2022,
	title = {Entangling single atoms over 33 km telecom fibre},
	volume = {607},
	issn = {0028-0836, 1476-4687},
	doi = {10.1038/s41586-022-04764-4},
	language = {en},
	number = {7917},
	journal = {Nature},
	author = {Van Leent, Tim and Bock, Matthias and Fertig, Florian and Garthoff, Robert and Eppelt, Sebastian and Zhou, Yiru and Malik, Pooja and Seubert, Matthias and Bauer, Tobias and Rosenfeld, Wenjamin and Zhang, Wei and Becher, Christoph and Weinfurter, Harald},
	month = jul,
	year = {2022},
	pages = {69--73},

}

@article{Legero_2003,
   title={Time-resolved two-photon quantum interference},
   volume={77},
   ISSN={1432-0649},
   DOI={10.1007/s00340-003-1337-x},
   number={8},
   journal={Appl. Phys. B},
   publisher={Springer Science and Business Media LLC},
   author={Legero, T. and Wilk, T. and Kuhn, A. and Rempe, G.},
   year={2003},
   month=Dec, pages={797–802} }

@article{Loredo:16,
author = {Juan C. Loredo and Nor A. Zakaria and Niccolo Somaschi and Carlos Anton and Lorenzo de Santis and Valerian Giesz and Thomas Grange and Matthew A. Broome and Olivier Gazzano and Guillaume Coppola and Isabelle Sagnes and Aristide Lemaitre and Alexia Auffeves and Pascale Senellart and Marcelo P. Almeida and Andrew G. White},
journal = {Optica},
number = {4},
pages = {433--440},
publisher = {Optica Publishing Group},
title = {Scalable performance in solid-state single-photon sources},
volume = {3},
month = {Apr},
year = {2016},
doi = {10.1364/OPTICA.3.000433},
}

@article{Barrett2005,
	title = {Efficient high-fidelity quantum computation using matter qubits and linear optics},
	volume = {71},
	doi = {10.1103/PhysRevA.71.060310},
	number = {6},
	journal = {Phys. Rev. A},
	publisher = {American Physical Society},
	author = {Barrett, Sean D. and Kok, Pieter},
	month = jun,
	year = {2005},
	pages = {060310},
}

@article{Pirandola2017,
  author       = {Pirandola, S. and Laurenza, R. and Ottaviani, C. and Banchi, L.},
  year         = {2017},
  journal = {Nat. Commun.},
  title        = {{Fundamental limits of repeaterless quantum communications}},
  doi          = {10.1038/ncomms15043},
  pages        = {15043},
  volume       = {8},
  }

@article{Bartling2025,
	title = {Universal high-fidelity quantum gates for spin qubits in diamond},
	volume = {23},
	doi = {10.1103/PhysRevApplied.23.034052},
	number = {3},
	journal = {Phys. Rev. Appl.},
	publisher = {American Physical Society},
	author = {Bartling, H.P. and Yun, J. and Schymik, K.N. and van Riggelen, M. and Enthoven, L.A. and van Ommen, H.B. and Babaie, M. and Sebastiano, F. and Markham, M. and Twitchen, D.J. and Taminiau, T.H.},
	month = mar,
	year = {2025},
	pages = {034052},

}

@article{Kuruma2021,
  author       = {Kuruma, Kazuhiro and Pingault, Benjamin and Chia, Cleaven and Renaud, Dylan and Hoffmann, Patrick and Iwamoto, Satoshi and Ronning, Carsten and Lončar, Marko},
  date         = {2021-06},
  journal = {Appl. Phys. Lett.},
	year		= {2021},
  title        = {Coupling of a single tin-vacancy center to a photonic crystal cavity in diamond},
  doi          = {10.1063/5.0051675},
  issn         = {1077-3118},
  number       = {23},
  volume       = {118},
	pages				= {230601},
  creationdate = {2026-05-28T10:22:44},
  publisher    = {AIP Publishing},
}

@article{Rugar2021,
	title = {Quantum Photonic Interface for {T}in-{V}acancy Centers in Diamond},
	volume = {11},
	doi = {10.1103/PhysRevX.11.031021},
	number = {3},
	journal = {Phys. Rev. X},
	publisher = {American Physical Society},
	author = {Rugar, Alison E. and Aghaeimeibodi, Shahriar and Riedel, Daniel and Dory, Constantin and Lu, Haiyu and McQuade, Patrick J. and Shen, Zhi-Xun and Melosh, Nicholas A. and Vučković, Jelena},
	month = jul,
	year = {2021},
	pages = {031021},
}

@article{Burek2017,
  author       = {Burek, Michael J. and Meuwly, Charles and Evans, Ruffin E. and Bhaskar, Mihir K. and Sipahigil, Alp and Meesala, Srujan and Machielse, Bartholomeus and Sukachev, Denis D. and Nguyen, Christian T. and Pacheco, Jose L. and Bielejec, Edward and Lukin, Mikhail D. and Lončar, Marko},
  date         = {2017-08},
  journal      = {Phys. Rev. Appl.},
  title        = {Fiber-Coupled Diamond Quantum Nanophotonic Interface},
  doi          = {10.1103/physrevapplied.8.024026},
  issn         = {2331-7019},
  number       = {2},
  pages        = {024026},
  year         = {2017},
  volume       = {8},
  creationdate = {2026-05-28T10:39:53},
  publisher    = {American Physical Society (APS)},
}

@article{Holl2025,
    author = {Holl, Peter and Adler, Steffen and Diwo-Emmer, Elke and Bächle, Andreas and Bradler, Maximilian and Yahyapour, Milad and Holzwarth, Ronald and Schäfer, Marlon and Becher, Christoph and Rattunde, Marcel},
    title = {2.5\,{W} {G}a{S}b-based {VECSEL} at 2062.4\,nm with an absolute wavelength stability below 1\,{MH}z},
    journal = {Appl. Phys. Lett.},
    volume = {127},
    number = {16},
    pages = {163302},
    year = {2025},
    month = {10},
    issn = {0003-6951},
    doi = {10.1063/5.0293416},
}

@article{kumar1990,
	title = {Quantum frequency conversion},
	copyright = {© 1990 Optical Society of America},
	doi = {10.1364/OL.15.001476},
	language = {EN},
	urldate = {2026-06-22},
	journal = {Opt. Lett.}, 
	volume = {15},
  pages = {1476},
	publisher = {Optica Publishing Group},
	author = {Kumar, Prem},
	month = dec,
	year = {1990},
    }

@article{Rohde2006,
	title = {Error models for mode mismatch in linear optics quantum computing},
	volume = {73},
	doi = {10.1103/PhysRevA.73.062312},
	number = {6},
	journal = {Phys. Rev. A},
	publisher = {American Physical Society},
	author = {Rohde, Peter P. and Ralph, Timothy C.},
	month = jun,
	year = {2006},
	pages = {062312},
}

@article{Arjona2022,
	title = {Photonic {Indistinguishability} of the {Tin}-{Vacancy} {Center} in {Nanostructured} {Diamond}},
	volume = {129},
	doi = {10.1103/PhysRevLett.129.173603},
	number = {17},
	journal = {Phys. Rev. Lett.},
	publisher = {American Physical Society},
	author = {Arjona Martínez, Jesús and Parker, Ryan A. and Chen, Kevin C. and Purser, Carola M. and Li, Linsen and Michaels, Cathryn P. and Stramma, Alexander M. and Debroux, Romain and Harris, Isaac B. and Hayhurst Appel, Martin and Nichols, Eleanor C. and Trusheim, Matthew E. and Gangloff, Dorian A. and Englund, Dirk and Atatüre, Mete},
	month = oct,
	year = {2022},
	pages = {173603},
}

@article{Brevoord2025,
	title = {Quantum frequency conversion of single photons from a tin-vacancy center in diamond},
	volume = {3},
	issn = {2837-6714},
	doi = {10.1364/OPTICAQ.576448},
	language = {EN},
	number = {6},
	journal = {Opt. Quantum},
	publisher = {Optica Publishing Group},
	author = {Brevoord, Julia Maria and Geus, Jan Fabian and Turan, Tim and Romero, Miguel Guerrero and Rodríguez, Daniel Bedialauneta and Codreanu, Nina and Stramma, Alexander Moritz and Hanson, Ronald and Elsen, Florian and Jungbluth, Bernd},
	month = dec,
	year = {2025},
	pages = {583--589},
}

@article{Stolk2024,
	title = {Metropolitan-scale heralded entanglement of solid-state qubits},
	volume = {10},
	doi = {10.1126/sciadv.adp6442},
	number = {44},
	journal = {Sci. Adv.},
	publisher = {American Association for the Advancement of Science},
	author = {Stolk, Arian J. and van der Enden, Kian L. and Slater, Marie-Christine and te Raa-Derckx, Ingmar and Botma, Pieter and van Rantwijk, Joris and Biemond, J. J. Benjamin and Hagen, Ronald A. J. and Herfst, Rodolf W. and Koek, Wouter D. and Meskers, Adrianus J. H. and Vollmer, René and van Zwet, Erwin J. and Markham, Matthew and Edmonds, Andrew M. and Geus, J. Fabian and Elsen, Florian and Jungbluth, Bernd and Haefner, Constantin and Tresp, Christoph and Stuhler, Jürgen and Ritter, Stephan and Hanson, Ronald},
	month = oct,
	year = {2024},
	pages = {eadp6442},
}

@unpublished{Waas2026,
  author = {Waas, Christopher and Dolné, Timo and Beukers, Hans K. C. and Stramma, Alexander M. and Codreanu, Nina and Mathieu, Noé and Hanson, Ronald},
  title = {Remote Entanglement of Solid-State Spin Qubits Integrated in Broadband Waveguides},
  note = {Preprint at https://arxiv.org/abs/2607.12002 (2026)}
}

@unpublished{Bushmakin2025,
  author = {Bushmakin, Vladislav and Berg, Oliver von and Sauerzapf, Colin and Jayaram, Sreehari and Denisenko, Andrej and Tarín, Cristina and Anders, Jens and Vorobyov, Vadim and Gerhardt, Ilja and Liu, Di and Wrachtrup, Jörg},
  title  = {Two-Photon Interference of Photons from Remote Tin-Vacancy Centers in Diamond},
  note   = {Preprint at https://arxiv.org/abs/2412.17539 (2025)}
}

@article{Wei2025,
	title = {Universal distributed blind quantum computing with solid-state qubits},
	volume = {388},
	doi = {10.1126/science.adu6894},
	number = {6746},
	journal = {Science},
	publisher = {American Association for the Advancement of Science},
	author = {Wei, Y.-C. and Stas, P.-J. and Suleymanzade, A. and Baranes, G. and Machado, F. and Huan, Y. Q. and Knaut, C. M. and Ding, S. W. and Merz, M. and Knall, E. N. and Yazlar, U. and Sirotin, M. and Wang, I. W. and Machielse, B. and Yelin, S. F. and Borregaard, J. and Park, H. and Lončar, M. and Lukin, M. D.},
	month = may,
	year = {2025},
	pages = {509--513},
}

@article{Stas2026,
	title = {Entanglement-assisted non-local optical interferometry in a quantum network},
	volume = {651},
	copyright = {2026 The Author(s)},
	issn = {1476-4687},
	doi = {10.1038/s41586-026-10171-w},
	language = {en},
	number = {8105},
	journal = {Nature},
	publisher = {Nature Publishing Group},
	author = {Stas, P.-J. and Wei, Y.-C. and Sirotin, M. and Huan, Y. Q. and Yazlar, U. and Abdo Arias, F. and Knyazev, E. and Baranes, G. and Machielse, B. and Grandi, S. and Riedel, D. and Borregaard, J. and Park, H. and Lončar, M. and Suleymanzade, A. and Lukin, M. D.},
	month = mar,
	year = {2026},
	pages = {326--332},
}

@article{Loredo2019,
	title = {Generation of non-classical light in a photon-number superposition},
	volume = {13},
	copyright = {2019 The Author(s), under exclusive licence to Springer Nature Limited},
	issn = {1749-4893},
	doi = {10.1038/s41566-019-0506-3},
	language = {en},
	number = {11},
	journal = {Nat. Photon.},
	publisher = {Nature Publishing Group},
	author = {Loredo, J. C. and Antón, C. and Reznychenko, B. and Hilaire, P. and Harouri, A. and Millet, C. and Ollivier, H. and Somaschi, N. and De Santis, L. and Lemaître, A. and Sagnes, I. and Lanco, L. and Auffèves, A. and Krebs, O. and Senellart, P.},
	month = nov,
	year = {2019},
	pages = {803--808},
}

@article{Thomas2014,
	title = {Silica based polishing of {100} and {111} single crystal diamond},
	doi = {10.1088/1468-6996/15/3/035013},
	journal = {Sci. Technol. Adv. Mater},
	volume = {15},
	number = {3},
	pages = {035013},
	author = {Thomas, Evan L. H. and Mandal, Soumen and Brousseau, Emmanuel B. and Williams, Oliver A.},
	month = jun,
	year = {2014},
}

@article{Weber2019,
	title = {Two-photon interference in the telecom {C}-band after frequency conversion of photons from remote quantum emitters},
	volume = {14},
	issn = {1748-3395},
	doi = {10.1038/s41565-018-0279-8},
	number = {1},
	journal = {Nat. Nanotechnol.},
	author = {Weber, Jonas H. and Kambs, Benjamin and Kettler, Jan and Kern, Simon and Maisch, Julian and Vural, Hüseyin and Jetter, Michael and Portalupi, Simone L. and Becher, Christoph and Michler, Peter},
	month = jan,
	year = {2019},
	pages = {23--26},
}
\onecolumn

\renewcommand{\figurename}{Extended Data Fig.} \setcounter{figure}{0}
\renewcommand{\tablename}{Extended Data Tab.} \setcounter{table}{0}
\begin{figure*}[htbp]
    \centering
    \begin{subfigure}[t]{0.48\textwidth}
        \centering
        \includegraphics[width=\linewidth]{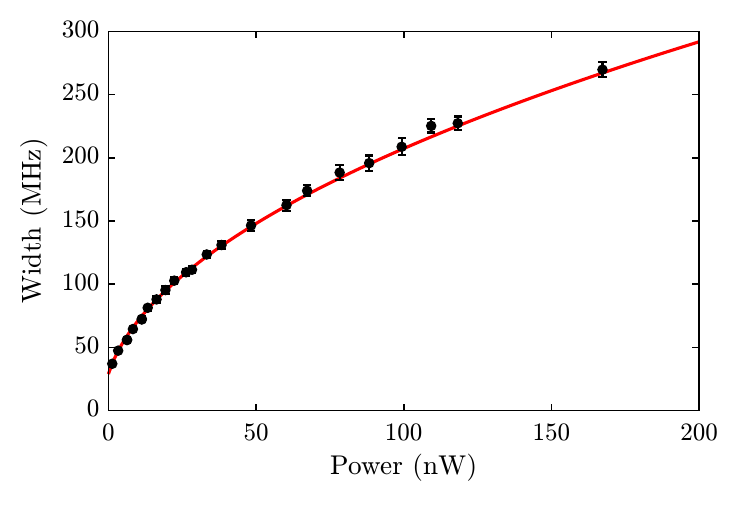} 
        \caption{}
        \label{EDF:Powerbroadening_A}
    \end{subfigure}\hfill
    \hfill
    \begin{subfigure}[t]{0.48\textwidth}
        \centering
        \includegraphics[width=\linewidth]{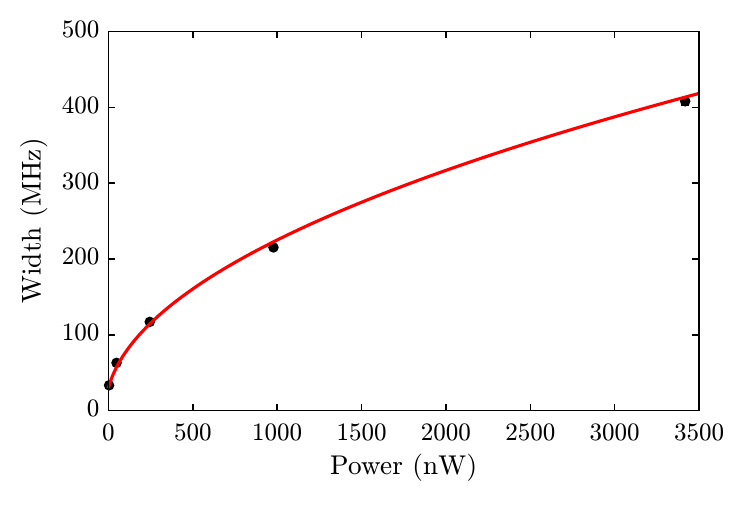} 
        \caption{}
        \label{EDF:Powerbroadening_B}
    \end{subfigure}\hfill

  \begin{subfigure}[t]{0.97\textwidth}
    \centering
    \includegraphics[width=\linewidth]{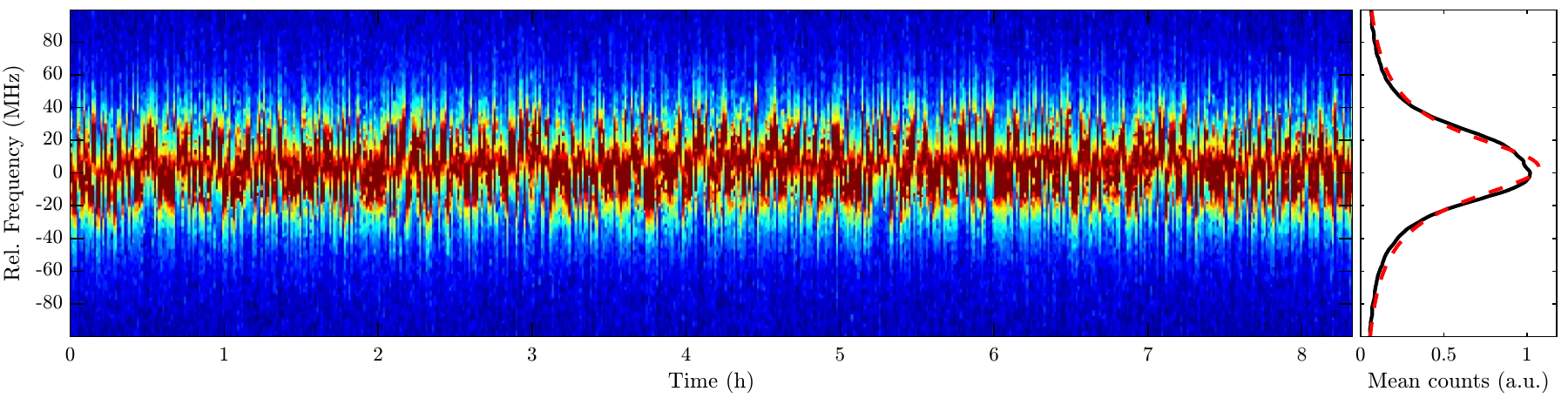} 
    \caption{}
     \label{EDF:PLE_RE1Longterm}
  \end{subfigure}
  \vskip\baselineskip 
\caption{\textbf{a}, Power broadening of the emission linewidth of a single SnV center in Sample A, measured by photoluminescence excitation (PLE) spectroscopy. A fit to the data (red curve) yields a zero-power linewidth of 29.0(3)\,MHz, in excellent agreement with the Fourier-transform-limited linewidth of 28.9(2)\,MHz. \textbf{b}, Corresponding measurement for the single SnV center in Sample Iso1 used for the HOM interference experiments. A fit to the data (red curve) yields a zero-power linewidth of 30(1)\,MHz, close to the Fourier-transform-limited value of 29.1(1)\,MHz, inferred from the measured excited-state lifetime of 5.46(2)\,ns. \textbf{c}, Long-term PLE spectrum of the same emitter as in \textbf{b}, recorded over 8\,h. The average of all scans (solid black line) is fitted with a Lorentzian function (red dashed line), yielding a integrated linewidth of 49.1(1)\,MHz. This broadening arises from residual power broadening and spectral jumps occurring over the measurement duration. For comparison, fitting an individual scan yields a linewidth of 34.1(2)\,MHz, expected from the power-broadening measurement.}
\end{figure*}

\begin{figure}[ht]
	\begin{subfigure}{0.48\textwidth}
         \includegraphics[width=\linewidth]{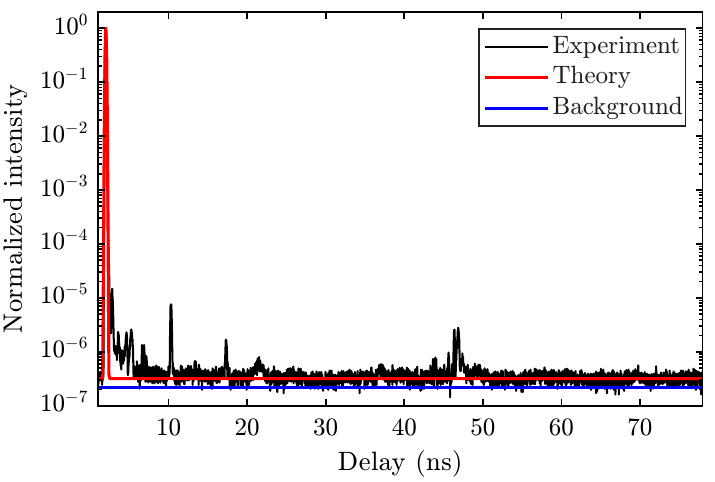} 
         \caption{} 
         \label{EDF:Pulses} 
	\end{subfigure}
	\hfill
	\begin{subfigure}{0.48\textwidth}
         \centering
        \includegraphics[width=\linewidth]{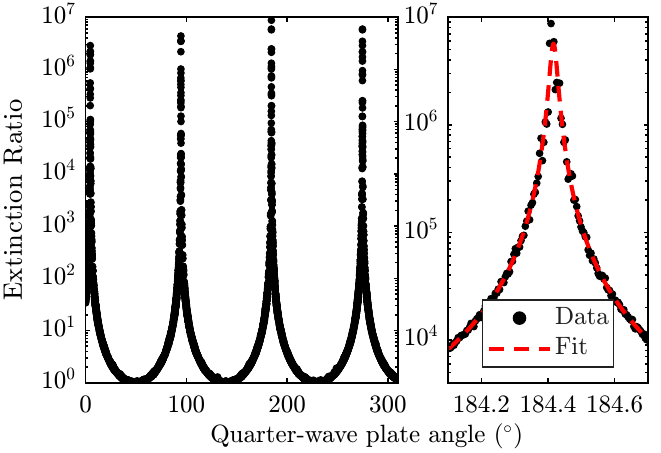} 
        \caption{}
         \label{EDF:ER_cross}
	\end{subfigure}
	\caption{\textbf{a}, TCSPC measurement of the pure excitation pulse recorded directly at a single SNSPD channel (black line). The red line represents the expected TCSPC curve for a gaussian pulse with an extinction ratio of $10^7$. The apparent extinction ratio is limited by the influence of the present dark count level (blue line). \textbf{b}, Measurement of the cross-polarization extinction depending on the angle of the quarter-wave plate. The best extinction ratio orientation is used for the experiment. From the fit we extract an extinction ratio of approximately $5.6\cdot10^6$.}
\end{figure}
   
\begin{table}[h]
\resizebox{\textwidth}{!}{%
\begin{tabular}{lllllllll}
		\toprule
			 &	$t^\mathrm{G2}$\,(ns)	&	Fraction\,(\%)  &	PM\,(°)	&	SD\,(MHz)	&	PD\,(MHz)	&	$V_{\mathrm{HOM}}^{\mathrm{raw}}$ &	$V_{\mathrm{HOM}}^{\mathrm{bc}}$	&	$M^{\mathrm{intr}}$  \\
		\midrule
			visible	&	38.75			&	93	&	$7.660^{+1.260}_{-3.020}$	&	$2.860^{+1.065}_{-2.860}$	&	$0.017^{+1.172}_{-0.017}$	&	$0.950^{+0.006}_{-0.008}$	&	$0.979^{+0.006}_{-0.009}$	&	$0.997^{+0.003}_{-0.011}$\\
			visible	&	7.3       		&	65	&	$7.886^{+1.346}_{-3.249}$	&	$5.233^{+5.672}_{-5.231}$	&	$0.000^{+2.812}_{-0.000}$	&	$0.974^{+0.007}_{-0.009}$	&	$0.980^{+0.007}_{-0.009}$	&	$0.999^{+0.001}_{-0.012}$\\
		\midrule
			telecom	&	38.75			&	93	&	$3.712^{+4.370}_{-3.711}$	&	$0.003^{+6.567}_{-0.003}$	&	$0.798^{+2.490}_{-0.797}$	&	$0.803^{+0.007}_{-0.028}$	&	$0.987^{+0.009}_{-0.034}$	&	$0.992^{+0.008}_{-0.031}$\\
			telecom	&	7.3			&	65    &	$4.441^{+3.966}_{-4.440}$	&	$0.017^{+16.304}_{-0.015}$	&	$0.000^{+5.203}_{-0.000}$	&	$0.940^{+0.004}_{-0.029}$	&	$0.994^{+0.005}_{-0.031}$	&	$0.999^{+0.001}_{-0.026}$\\
		\bottomrule
	\end{tabular}%
    }
    \caption{Fitting results with 95\% credible interval. In our fitting routine, spectral diffusion (SD) and pure dephasing (PD) of the emitter, as well as polarization mismatch (PM) introduced by the MZI, constitute the only free parameters, while other quantities, such as SBR, are determined independently.}
    \label{EDF:Fit_results}
\end{table}

\end{document}